\documentclass[a4paper,12pt,preprint,amsmath,showpacs,
nofootinbib,superscriptaddress]{revtex4}
\usepackage{mathrsfs}
\usepackage{amssymb}
\usepackage{amsmath}
\usepackage{graphicx}
\usepackage{multirow}
\usepackage{subfigure}
\usepackage{float}
\usepackage{hyperref}
\usepackage{graphics}

\def\as{\alpha_s}

\def\nno{\nonumber}

\begin{document}
\title{Transverse-Momentum Resummation for Gauge Boson Pair Production at the Hadron Collider}
\author{Yan  Wang}
\affiliation{Department of Physics and State Key Laboratory of
Nuclear Physics and Technology, Peking University, Beijing 100871,
China}
\author{Chong Sheng Li}
\email{csli@pku.edu.cn}
\affiliation{Department of Physics and State Key Laboratory of
Nuclear Physics and Technology, Peking University, Beijing 100871,
China}
\affiliation{Center for High Energy Physics, Peking University, Beijing, 100871, China}
\author{Ze Long Liu}
\affiliation{Department of Physics and State Key Laboratory of
Nuclear Physics and Technology, Peking University, Beijing 100871,
China}
\author{Ding Yu Shao}
\affiliation{Department of Physics and State Key Laboratory of
Nuclear Physics and Technology, Peking University, Beijing 100871,
China}
\author{Hai Tao Li}
\affiliation{Department of Physics and State Key Laboratory of
Nuclear Physics and Technology, Peking University, Beijing 100871,
China}



\begin{abstract}
We perform  the  transverse-momentum resummation for $W^{+}W^{-}$, $ZZ$, and $W^{\pm}Z$ pair productions at the  next-to-next-to-leading logarithmic accuracy using soft-collinear effective theory for $\sqrt{S}=8~\text{TeV}$ and $\sqrt{S}=14~\text{TeV}$  at the LHC, respectively. Especially, this is the first calculation of $W^{\pm}Z$ transverse-momentum resummation. We also include the non-perturbative effects and discussions on the PDF uncertainties. Comparing with the next-to-leading logarithmic results, the next-to-next-to-leading logarithmic resummation can reduce the dependence of the transverse-momentum distribution on the factorization scales significantly. Finally, we find that our numerical results are consistent with data measured by CMS collaboration for the $ZZ$ production, which have been only reported by the LHC experiments for the unfolded transverse-momentum distribution of the gauge boson pair production so far, within theoretical and experimental uncertainties.
\end{abstract}

\maketitle

\section{Introduction}\label{s1}
The gauge boson pair productions are important within and beyond the Standard Model (SM). For the cases of $W^+W^-$ and $W^{\pm}Z$ productions, they can be used to test the non-Abelian gauge structure, especially triple-gauge-boson couplings. Besides,  $W^+W^-$ and $ZZ$ are irreducible SM backgrounds of Higgs boson production. If there is any deviation from the predictions of SM, it may be a new physics signal. Therefore, it is essential to count on accurate theoretical predictions for these processes.

Experimental collaborations at the Tevatron and the LHC have reported experimental results of various kinematic distributions for $W^+W^-$, $ZZ$, $W^{\pm}Z$ productions. The leptonic decay mode of gauge boson pair has been analyzed at the Tevatron~\cite{Aaltonen:2009aa,CDF:2011ab,Abazov:2013opa,Abazov:2012cj,D0:2013rca}, and at the LHC for $\sqrt{S}=$7 TeV and  $\sqrt{S}=$8 TeV, respectively~\cite{Aad:2011xj,Chatrchyan:2012sga}. Especially, more stringent limitations on anomalous triple-gauge-boson couplings than in the past have been presented by the LHC collaborations ~\cite{Aad:2011kk,Aad:2012twa,Aad:2011cx}.

Furthermore, if the gauge boson pair comes from the decay of a heavy resonance,  the kinematics of the gauge boson pair will carry information of the resonance. Therefore, it is necessary to consider the boson pair as a unit, rather than each individual gauge boson. The transverse-momentum $q_T$ of the boson pair system is one important observable, which has been measured at the LHC~\cite{ATLAS:2012mec,Aad:2012twa,CMS-PAS-SMP-13-005}.

The next-to-leading order (NLO) QCD corrections to $W^+W^-$, $ZZ$ and $W^{\pm}Z$ production were calculated many years ago~\cite{Frixione:1993yp,PhysRevD.43.3626,Mele1991409,PhysRevD.60.114037,PhysRevD.60.113006}. Besides, NLO QCD corrections with helicity amplitudes method were completed in Ref.~\cite{Dixon19983}, where the effects of spin correlation were fully taken into account. Recently, two-loop virtual QCD corrections to $W^+W^-$  production in the high energy limit have been reported in Ref.~\cite{Chachamis2008a}, and threshold resummation in the soft-collinear effective theory (SCET) and  the approximate NNLO cross sections for $W^+W^-$ production are calculated in Ref.~\cite{Dawson:2013lya}. And $W^{\pm}Z$ production is calculated beyond NLO QCD for high $q_T$ region~\cite{Campanario:2012fk}.   However, when the invariant mass of gauge boson pair $M$ is much larger than $q_T$, there exists large logarithmic terms of the form $\ln(q_T^2/M^2)$ in the small $q_T$ region. The fixed-order predictions are invalid in this region. Therefore it is necessary to resum these large logarithmic terms to all order.

In this paper, we calculate the transverse-momentum resummation of the gauge boson pair production at the next-to-next-leading logarithmic (NNLL) accuracy based on SCET~\cite{Bauer2001,Bauer2002,Beneke2002}. In the case of transverse-momentum resummation, frameworks equivalent to the Collins, Soper and Sterman (CSS) formalism have been developed for both the Drell-Yan process and Higgs production~\cite{Gao:2005iu,Idilbi:2005er,Becher:2010tm,Becher:2011xn,GarciaEchevarria:2011rb,Chiu:2012ir,Becher:2012yn,Becher2011,Becher2012}.
The framework we adopted in the paper is built upon Refs.~\cite{Becher2011,Becher2012}. In the case of $W^+W^-$ and $ZZ$ pair production, $q_T$ resummation has been discussed  in the CSS framework~\cite{Grazzini2005a,Frederix2008,Balazs1999,Balazs2001a}.
However, to our knowledge, the resummation  effects on the transverse-momentum of $W^{\pm}Z$ production have not been calculated so far.

The paper is organized as follows. In Sec.~\ref{s2}, we describe the formalism for $q_T$ resummation in SCET briefly. In Sec.~\ref{s3}, we present our numerical results. Then section~\ref{s4} is a brief  conclusion.

\section{Factorization and Resummation}\label{s2}
In this section, we briefly review the transverse-momentum resummation in SCET formalism in Refs.~\cite{Becher2011,Becher2012}. The resummation formulas of transverse-momentum distribution we used can be applied to the processes where non-strongly interacting particles are produced in hadronic collisions.

We consider the processes
\begin{eqnarray}\label{s2_eq_process}
N_1(P_1) + N_2(P_2) \rightarrow V_l(p_3) + V_m(p_4) + X(p_x),
\end{eqnarray}
where $V_{l,m}~(l,m=W, Z)$ is  $W$ or $Z$ boson, and $X$ is an inclusive hadronic final state. In the Born level, the partonic process is
\begin{eqnarray}\label{s2_eq_parton_process}
q(p_1) + q'(p_2) \rightarrow V_l(p_3) + V_m(p_4),
\end{eqnarray}
where $p_{i} = z_{i} P_i,~ i=1,2$. The kinematic variables are defined as follows
\begin{eqnarray}\label{s2_eq_kinematic_variables}
&&S = (P_1 + P_2)^2,\quad  s = (p_1 + p_2)^2, \quad t=(p_1-p_3)^2, \quad u=(p_2-p_3)^2, \nonumber \\
&& q=p_3 + p_4, \quad q^2= M^2,\quad \tau = (M^2+q_T^2)/S.
\end{eqnarray}
In the kinematical region of $\Lambda_{QCD}^2\ll q_T^2 \ll M$, soft and collinear emissions  can be treated in the SCET frame. The gauge boson pair differential cross section can be factorized  as follows~\cite{Li:2013mia}:
\begin{eqnarray}\label{s2_eq_sigma_OPE}
\frac{d^3\sigma}{dq_T^2dydM^2} &=& \frac{1}{S}  \mathcal{H}_{V_{l}V_{m}}(M,\mu)  \frac{1}{4\pi}\int d^2x_{\perp} e^{-i q_{\perp}\cdot x_{\perp}}  \nonumber\\
&&\times \sum_{q,q'} \big[\mathcal{B}_{q/N_1}\left(z_1,x_T^2,\mu_f \right)\mathcal{B}_{q'/N_2}\left(z_2,x_T^2,\mu_f \right) + (q\leftrightarrow q')\big],
\end{eqnarray}
where $y$ is the rapidity of the boson pair system, $\mu_f$ is the factorization scale and $z_{1,2}=\sqrt{\tau} e^{\pm y}$. Here $\mathcal{H}_{V_{l}V_{m}}$ is the hard function and can be expanded as
\begin{eqnarray}
 \mathcal{H}_{V_{l}V_{m}}= \mathcal{H}_{V_{l}V_{m}}^{(0)} + \frac{\alpha_s}{4\pi}\mathcal{H}_{V_{l}V_{m}}^{(1)} + \cdots .
\end{eqnarray}
As a cross-check, we recalculate $\mathcal{H}_{V_{l}V_{m}}^{(0)}$ and $\mathcal{H}_{V_{l}V_{m}}^{(1)}$ and find that our results are consistent with those in Refs.~\cite{Frixione:1993yp,Frixione:1992pj,PhysRevD.43.3626}, the corresponding details are listed in the App.~\ref{appa}. The RG equation of hard function can be written as
\begin{eqnarray}
\frac{d}{d\ln\mu} \mathcal{H}_{V_{l}V_{m}}\left(M,\mu_f \right)= 2\left[ \Gamma_{\rm cusp}^F(\alpha_s)\ln\frac{-M^2}{\mu_f^2} + 2\gamma^{V}(\alpha_s)\right]\mathcal{H}_{V_{l}V_{m}}\left(M,\mu_f \right),
\end{eqnarray}
where $\Gamma_{\rm cusp}^F(\alpha_s)$ is the cusp anomalous dimension of Wilson loops with light-like segments, while $\gamma^V(\alpha_s)$ controls the single-logarithmic evolution. After solving the RG equation, we obtain the hard function
\begin{align}
\mathcal{H}_{V_{l}V_{m}}\left(M,\mu_f \right) &= \nno\\
 &\hspace{-4em} \exp\left[ 4S(\mu_h^2,\mu_f^2) - 2a_{\Gamma}(\mu_h^2,\mu_f^2)\ln\frac{-M^2}{\mu_h^2} - 4a_{\gamma^V}(\mu_h^2,\mu_f^2) \right] \mathcal{H}_{V_{l}V_{m}}\left(M,\mu_h \right),
\end{align}
where $\mu_h$ is the hard matching scale. Here $S(\nu^2,\mu^2)$ and $a_\Gamma(\nu^2,\mu^2)$ are defined as
\begin{eqnarray}
 S(\nu^2,\mu^2)&=&-\int_{\as(\nu^2)}^{\as(\mu^2)}d\alpha \frac{\Gamma_{\rm cusp}^F(\alpha)}{\beta(\alpha)}
 \int_{\as(\nu^2)}^{\alpha}\frac{d\alpha'}{\beta(\alpha')}, \\
 a_\Gamma(\nu^2,\mu^2)&=&-\int_{\as(\nu^2)}^{\as(\mu^2)}d\alpha \frac{\Gamma_{\rm cusp}^F(\alpha)}{\beta(\alpha)}.
\end{eqnarray}
$a_{\gamma^V}$ has a similar expression. Up to NNLL level, we need $3$-loop cusp anomalous dimension and $2$-loop normal anomalous dimension, and their explicit expressions are collected in the Appendices of Refs.~\cite{Becher:2007ty}.

The function $\mathcal{B}_{q/N}$ in Eq.~\ref{s2_eq_sigma_OPE} is the transverse-momentum dependent PDFs, which is defined by operator product expansion~\cite{Becher2011}. We adopt the analytic regularization of Ref.~\cite{Becher:2011dz}, and the product of the two $\mathcal{B}_{q/N}$ can be re-factorized as
\begin{align}
\left[ \mathcal{B}_{q/N_1}\left(z_1,x_T^2,\mu_f \right)\mathcal{B}_{q'/N_2}\left(z_2,x_T^2,\mu_f \right) \right]_{q^2} = & \nonumber \\
 &\hspace{-6em}  \left(\frac{x_T^2q_T^2}{4e^{-2\gamma_E}} \right)^{-F_{qq'}\left(x_T^2,\mu_f \right)} B_{q/N_1}\left(z_1,x_T^2,\mu_f \right)
   B_{q'/N_2}\left(z_2,x_T^2,\mu_f \right),
\end{align}
where $F_{qq'}$ controls hidden $q_T^2$ dependence induced by collinear anomaly~\cite{Becher2011} and $B_{q/N}\left(z,x_T^2,\mu_f \right)$ can be matched onto the standard PDF via:
\begin{eqnarray}
B_{q/N}\left(z,x_T^2,\mu_f \right)=\sum_i \int  I_{q\leftarrow i}\left(\xi,x_T^2,\mu_f
   \right) \phi
   _{i/N}\left(z/\xi,\mu_f \right) \frac{d\xi}{\xi}  + \mathcal{O}\left(\Lambda _{\text{QCD}}^2 x_T^2\right),
\end{eqnarray}
where $x_T \ll \Lambda_{QCD}^{-1}$ and $I_{q\leftarrow i}\left(z,x_T^2,\mu_f \right)$ is the matching coefficient functions~\cite{Becher2012}.
The RG equations for the matching coefficient $I_{q\leftarrow i}\left(z,x_T^2,\mu_f\right)$ are given by
\begin{eqnarray}
\frac{d}{d\ln\mu}I_{q\leftarrow i}\left(z,x_T^2,\mu\right) &=& \left[\Gamma_{\rm cusp}^F(\alpha_s)L_{\perp} -2\gamma^q(\alpha_s) \right] I_{q\leftarrow i}\left(z,x_T^2,\mu\right) \nonumber \\
&& -\sum_j \int_z^1 \frac{d u}{u} I_{q\leftarrow j}\left(u,x_T^2,\mu\right)\mathcal{P}_{j\leftarrow i}(z/u, \alpha_s),
\end{eqnarray}
where $\mathcal{P}_{j\leftarrow i}$ is the DGLAP splitting functions and  $L_{\perp}$ is defined as
\begin{eqnarray}
L_{\perp}=\text{ln}\frac{x_T^2 \mu^2}{4e^{-2\gamma_E}}.
\end{eqnarray}
After factoring out the double logarithmic terms in $I_{q\leftarrow j}\left(z,x_T^2,\mu_f\right)$ we have
\begin{eqnarray}
I_{i\leftarrow j}\left(z,x_T^2,\mu_f\right)\equiv e^{h_F(L_{\perp},\alpha_s)} \bar{I}_{i\leftarrow j}(z,L_{\perp},\alpha_s),
\end{eqnarray}
where $\bar{I}_{i \leftarrow j}$ satisfies as DGLAP equation with an opposite sign~\cite{Becher2012}, and the RG equation for $h_F(L_{\perp},\alpha_s)$ is
\begin{eqnarray}
 \frac{d}{d \ln\mu} e^{h_F(L_{\perp},\alpha_s)} = \Gamma_{\rm cusp}^{F}(\alpha_s) L_{\perp} -2 \gamma^q(\alpha_s).
\end{eqnarray}
After combining above results, we can get the factorized cross section
\begin{eqnarray}\label{s2_eq_sigma_final}
\frac{d^2\sigma}{d q_T^2 dy } &=& \frac{1}{S}  \sum_{i,j=q,q',g} \mathcal{H}_{VV}(M,\mu_f)\int_{\xi_1}^1 \frac{dz_1}{z_1} \int_{\xi_2}^1 \frac{dz_2}{z_2} \bar{C}_{qq'\rightarrow ij}\left(z_1,z_2,q_T^2,\mu_f\right)
\nonumber\\
&~&\times\phi_{i/N_1}(\xi_1/z_1,\mu_f) \phi_{j/N_2}(\xi_2/z_2,\mu_f) + (q,i\leftrightarrow q',j)\big].
\end{eqnarray}
Here $\bar{C}_{qq'\rightarrow ij}$ is the hard kernel of the process and defined as
\begin{eqnarray}\label{s2_eq_C}
\bar{C}_{qq'\rightarrow ij}\left(z_1,z_2,q_T^2,\mu_f \right) &=& \frac{1}{2} \int_0^{\infty}dx_T x_T J_0(x_T q_T)~\exp\left[g_F(\mu_f,L_{\perp},\alpha_s)\right] \nonumber\\
&&\times \left[\bar{I}_{q\leftarrow i}(z_1,L_{\perp},\alpha_s) \bar{I}_{q'\leftarrow j}(z_2,L_{\perp},\alpha_s)\right],
\end{eqnarray}
where $J_0$ is the zeroth order Bessel function, and $g_F(\eta,L_{\perp},\alpha_s)$ combines all the exponent terms~\cite{Becher2012}.

In addition to singular terms, which are resummed by Eq.~(\ref{s2_eq_sigma_final}), fixed-order computation also contributes non-singular terms to the total cross section. We need to combine the resummation result and the fixed-order result together for the $q_T$ spectrum. Finally, in order to avoid double counting, the RG improved predictions for the transverse-momentum of the gauge boson pair can be written as~\cite{Becher2012}
\begin{eqnarray}\label{s2_eq_fullcs}
\frac{d\sigma^{\rm NNLL + NLO}}{dq_T} = \frac{d\sigma^{\rm NNLL}}{dq_T} +\left[\frac{d\sigma^{\rm NLO}}{dq_T}-\frac{d\sigma^{\rm NNLL}}{dq_T}\right]_{\rm expanded~to~NLO}.
\end{eqnarray}

\section{Numerical Results}\label{s3}
In this section, we present the numerical results for the transverse-momentum resummation effects on gauge boson pair productions at the LHC. Unless specified otherwise, we choose SM input parameters as~\cite{Beringer:1900zz}:
\begin{eqnarray}\label{sm_para}
 &&  m_W= 80.4 \textrm{~GeV}, \quad    m_Z = 91.19 \textrm{~GeV}, \quad \alpha(m_Z)=1/132.338.
\end{eqnarray}
We use the MSTW2008NNLO PDF set and the corresponding running QCD coupling constant. The QCD coupling constant has a flavor threshold at $\mu_b=4.75~\text{GeV}$ for the b quark.  The NLO QCD corrections in Eq.~(\ref{s2_eq_fullcs}) are calculated by MCFM~\cite{Campbell2000}.  The factorization scale is set as $\mu_f=q^*+q_T$~\cite{Becher2012}, and $q^*$ is defined as $q^*=M\exp(-2\pi/(\Gamma_0^F\alpha_s(q^*)))$. The default hard scale is chosen as $\mu_h=M$. The large logarithmic terms between hard scale and factorization scale are resumed by RG equations.

Note that since for $M\geq m_{V_{l}}+m_{V_{m}} \geq 160.8~\text{GeV}$, $q^*\geq 2.2$~GeV, which are larger than those in Drell-Yan process, where $q^* =1.88$~GeV. We therefore expect that the non-perturbative effects are smaller than those in Drell-Yan production.
\begin{figure}[h]
\begin{minipage}[t]{0.45\linewidth}
\centering
  \includegraphics[width=1.0\linewidth]{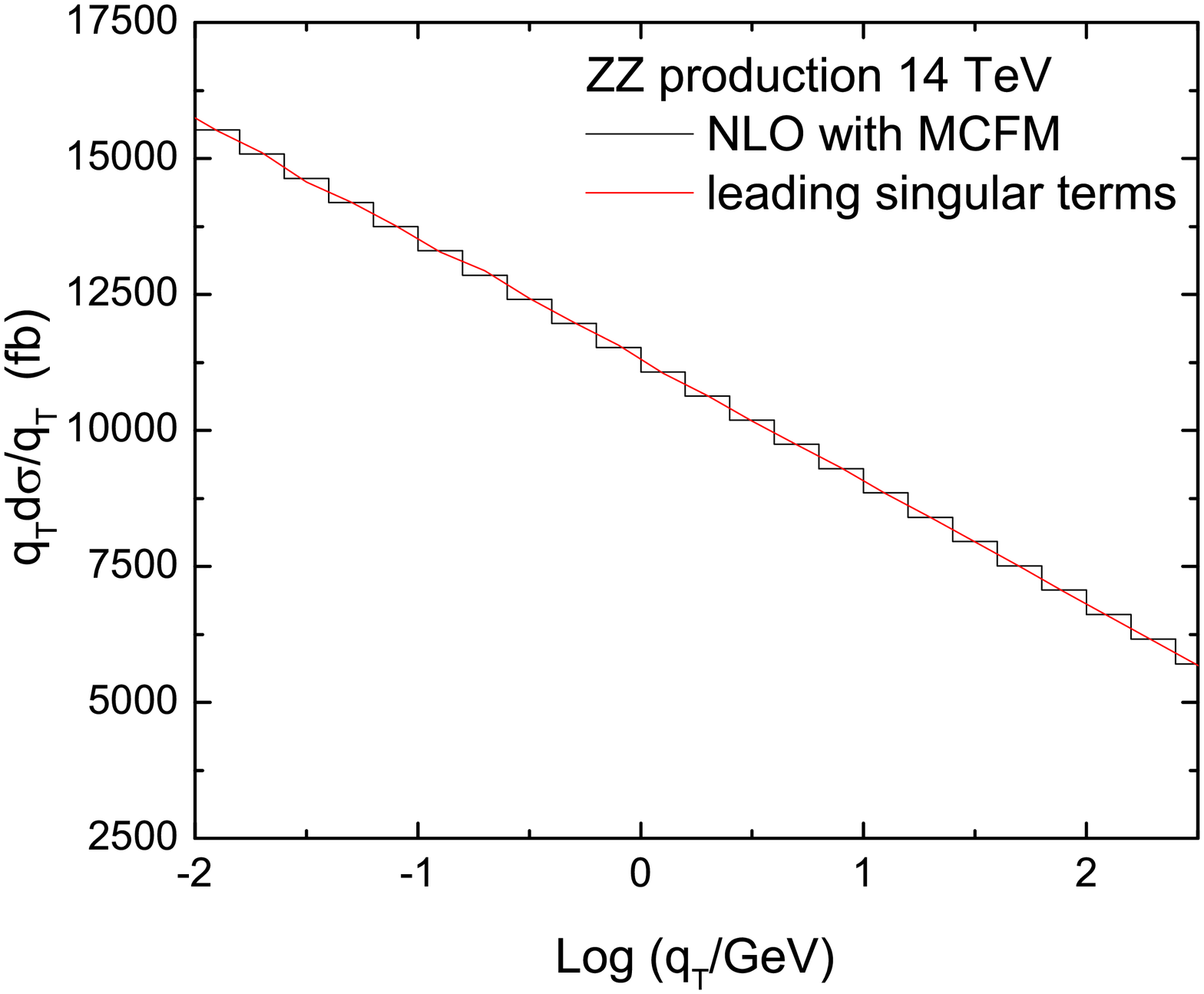}\\
\end{minipage}
\hfill
\begin{minipage}[t]{0.45\linewidth}
\centering
 \includegraphics[width=1.0\linewidth]{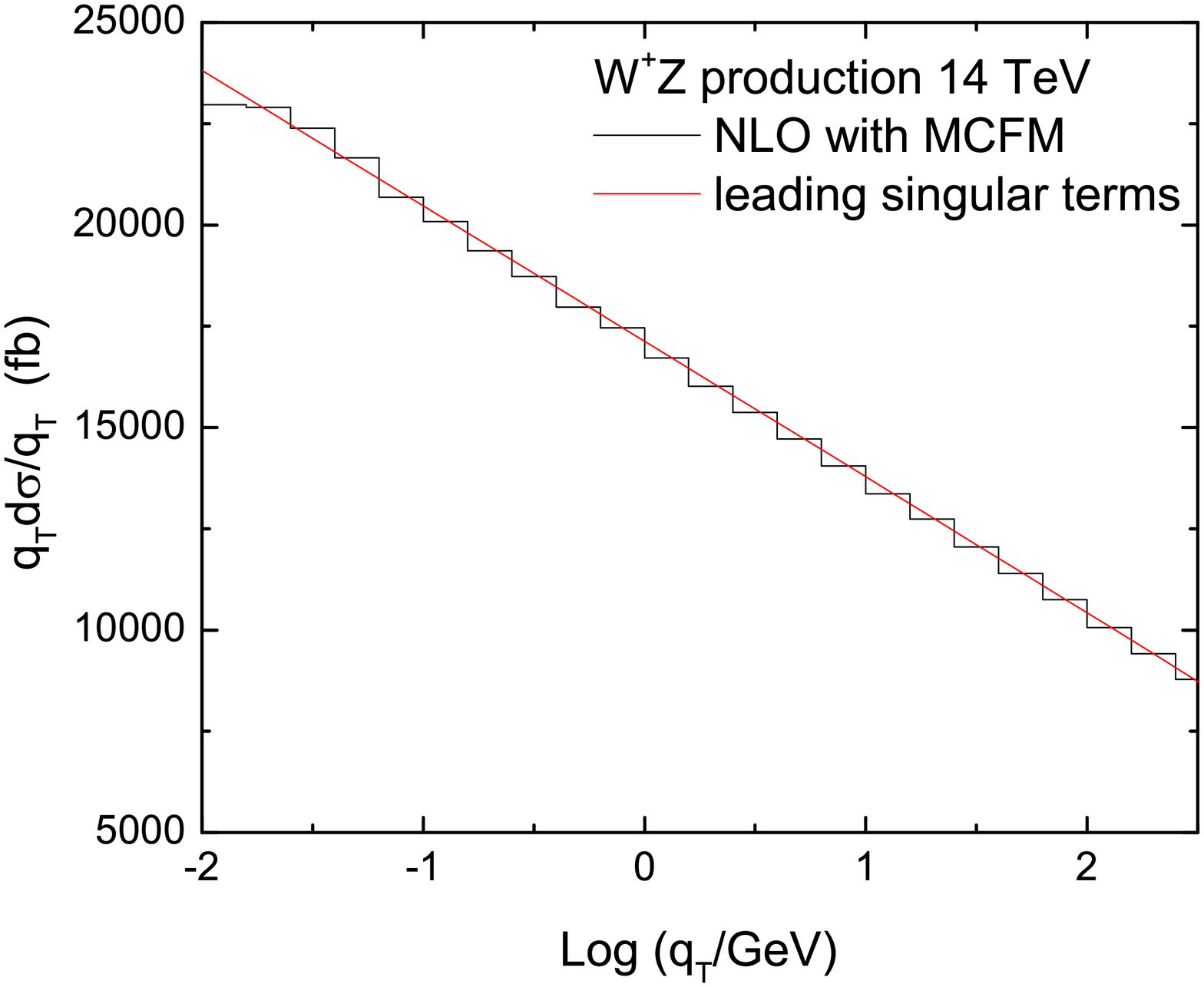}\\
\end{minipage}
\hfill
\begin{minipage}[t]{0.45\linewidth}
\centering
 \includegraphics[width=1.0\linewidth]{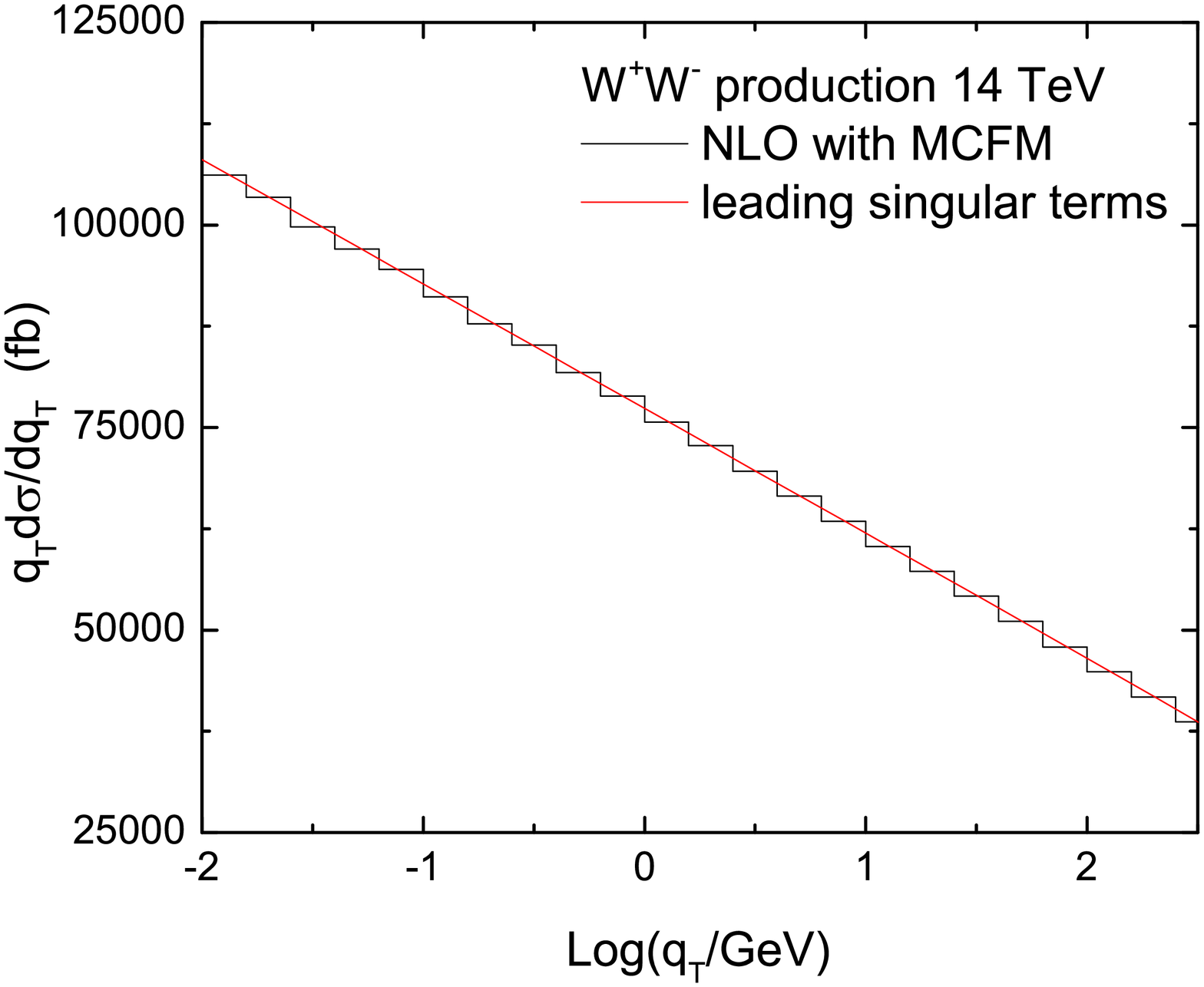}\\
\end{minipage}
\caption{\label{f_match}Comparison of the leading singular (red) and the exact NLO (black) distributions in the small $q_T$ region. }
\end{figure}

\subsection{Fixed-order Results And Non-perturbative Effects}
When resummation formula Eq.~(\ref{s2_eq_sigma_final}) is expanded to $\mathcal{O}(\alpha_s)$ in the limit $q_T\rightarrow 0$, the leading singular predictions should agree with the exact NLO results~\cite{Becher2012}. In Fig.~\ref{f_match}, we compare the leading singular results and exact NLO results calculated by MCFM. It is shown that they are consistent with each other.

\begin{figure}[t!]
\begin{minipage}[t]{0.45\linewidth}
\centering
  \includegraphics[width=1.0\linewidth]{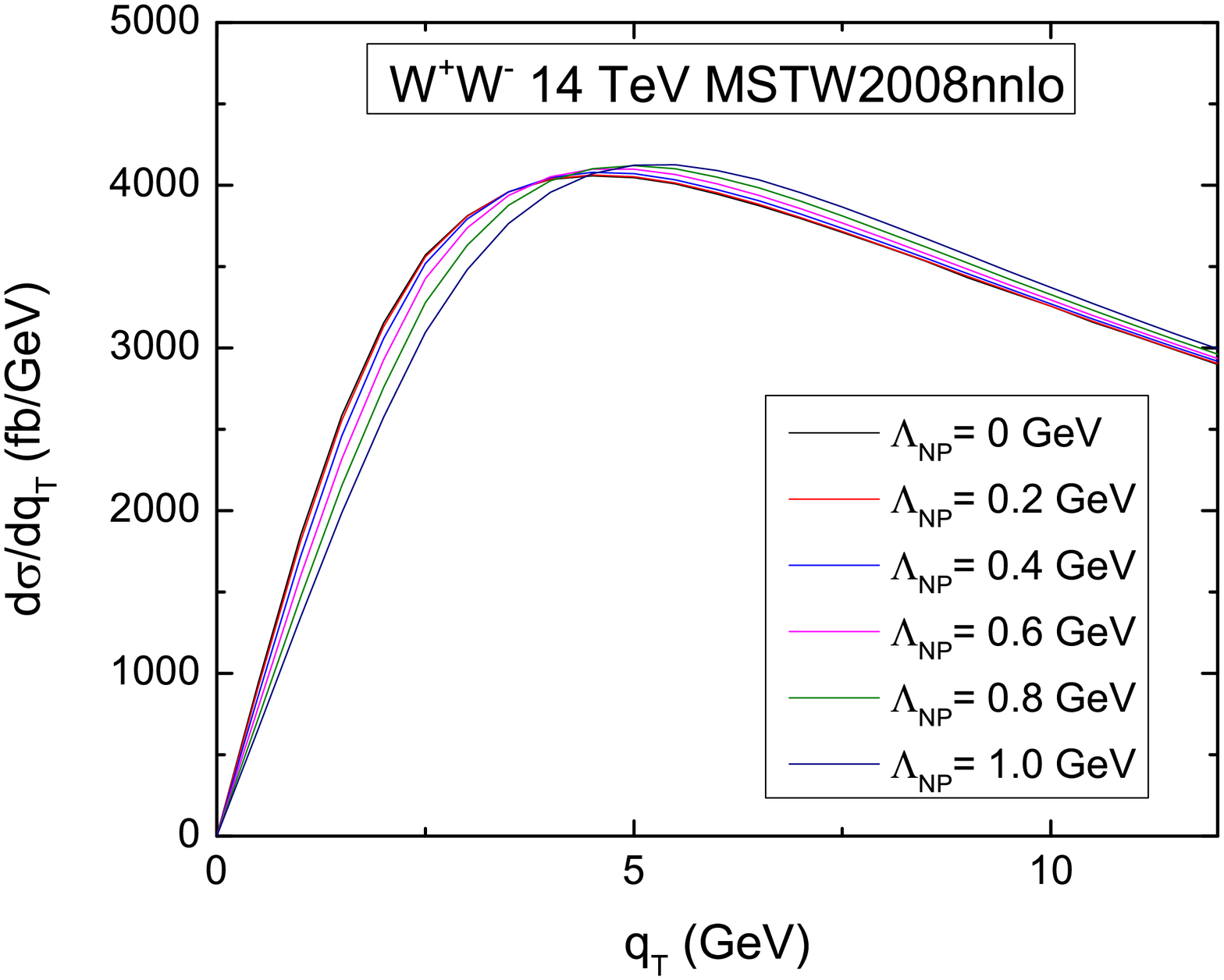}\\
\end{minipage}
\hfill
\begin{minipage}[t]{0.45\linewidth}
\centering
 \includegraphics[width=1.0\linewidth]{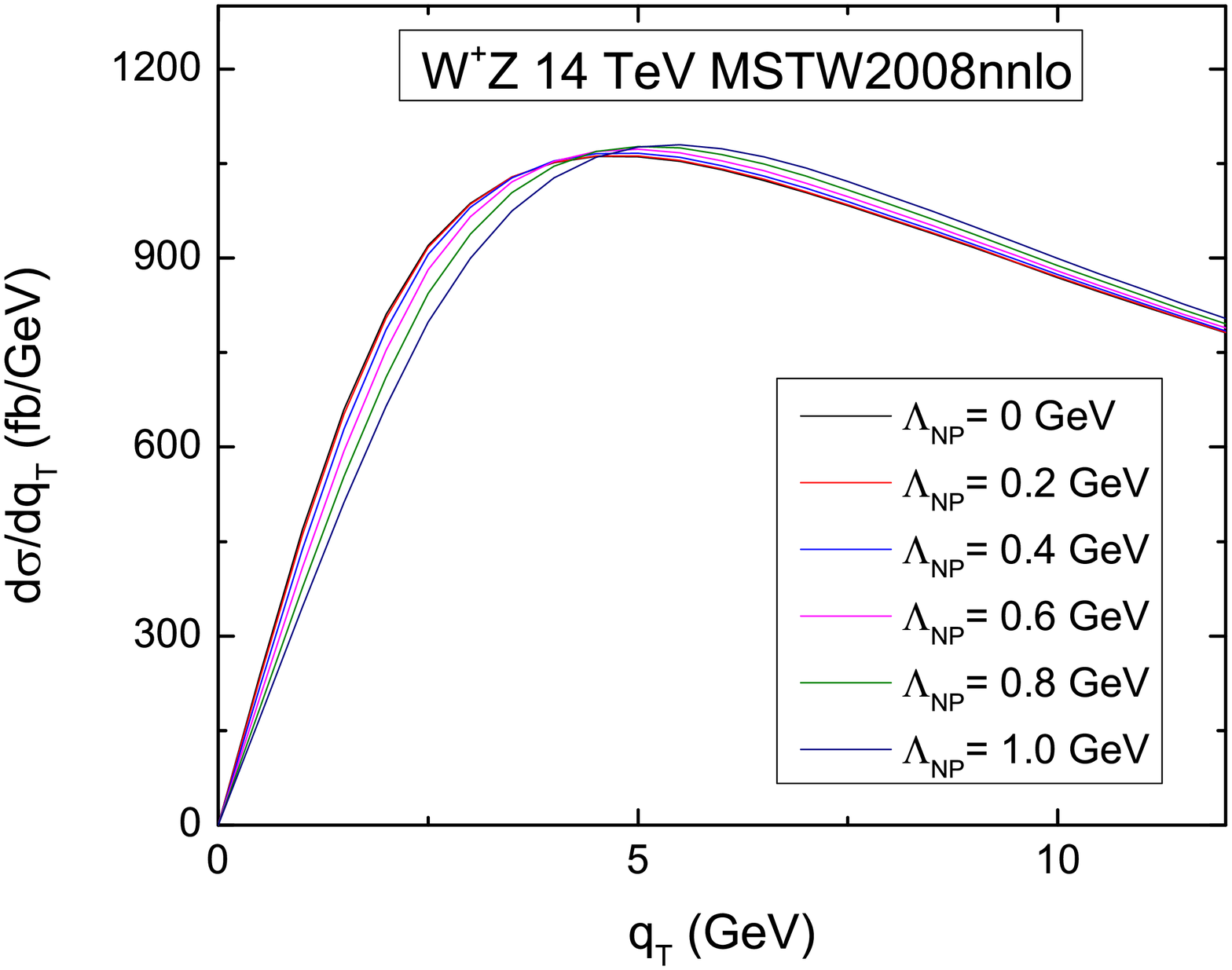}\\
\end{minipage}
\hfill
\begin{minipage}[t]{0.45\linewidth}
\centering
 \includegraphics[width=1.0\linewidth]{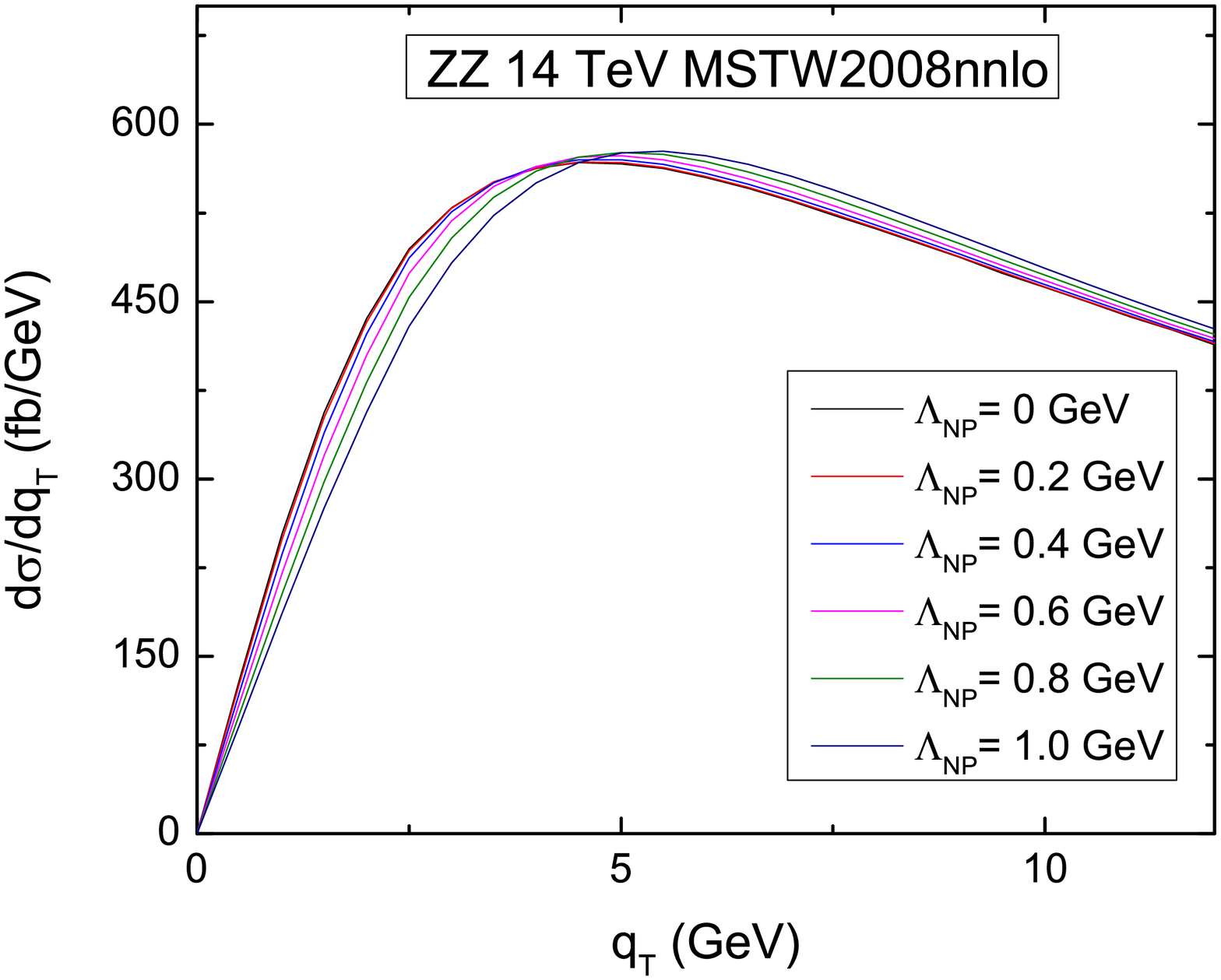}\\
\end{minipage}
\caption{\label{f_lp}The non-perturbative effects on the differential cross sections of gauge boson pair production with $\sqrt{S}=14~\text{TeV}$, respectively.}
\end{figure}

When discussing operator-product expansion of the transverse-position dependent PDFs $\mathcal{B}_{q/N}$,  a hadronic form factor $f_{\rm hadr}(x_T \Lambda_{NP})$, parameterized in terms of a hadronic scale $\Lambda_{NP}$, is needed to be introduced  in the region $q_T \sim \Lambda_{QCD}$~\cite{Becher2012}, and $\mathcal{B}_{q/N}$ can be expressed as
\begin{eqnarray}\label{s2_eq_le_beam}
\mathcal{B}_{q/N}(z,x_T^2,\mu_f) = \mathcal{B}_{q/N}^{pert}(z,x_T^2,\mu_f) f_{\rm hadr}(x_T \Lambda_{NP}).
\end{eqnarray}
Here the form factor has the form
\begin{eqnarray}\label{s2_eq_le}
f_{\rm hadr}(x_T \Lambda_{NP}) = \exp(-x_T^2 \Lambda_{NP}^2).
\end{eqnarray}
In Fig.~\ref{f_lp}, we present the non-perturbative effects on the differential cross sections of gauge boson pair resummation at the LHC with $\sqrt{S}=14~\text{TeV}$. Obviously, Fig.~\ref{f_lp} shows that the non-perturbative effects result in a tiny shift on the $q_T$ spectrum. We choose $\Lambda_{NP}=0.6$~GeV in the following numerical calculations.

\subsection{Resummation Results}
\begin{figure}[t!]
\begin{minipage}[t!]{0.45\linewidth}
\centering
  \includegraphics[width=1.0\linewidth]{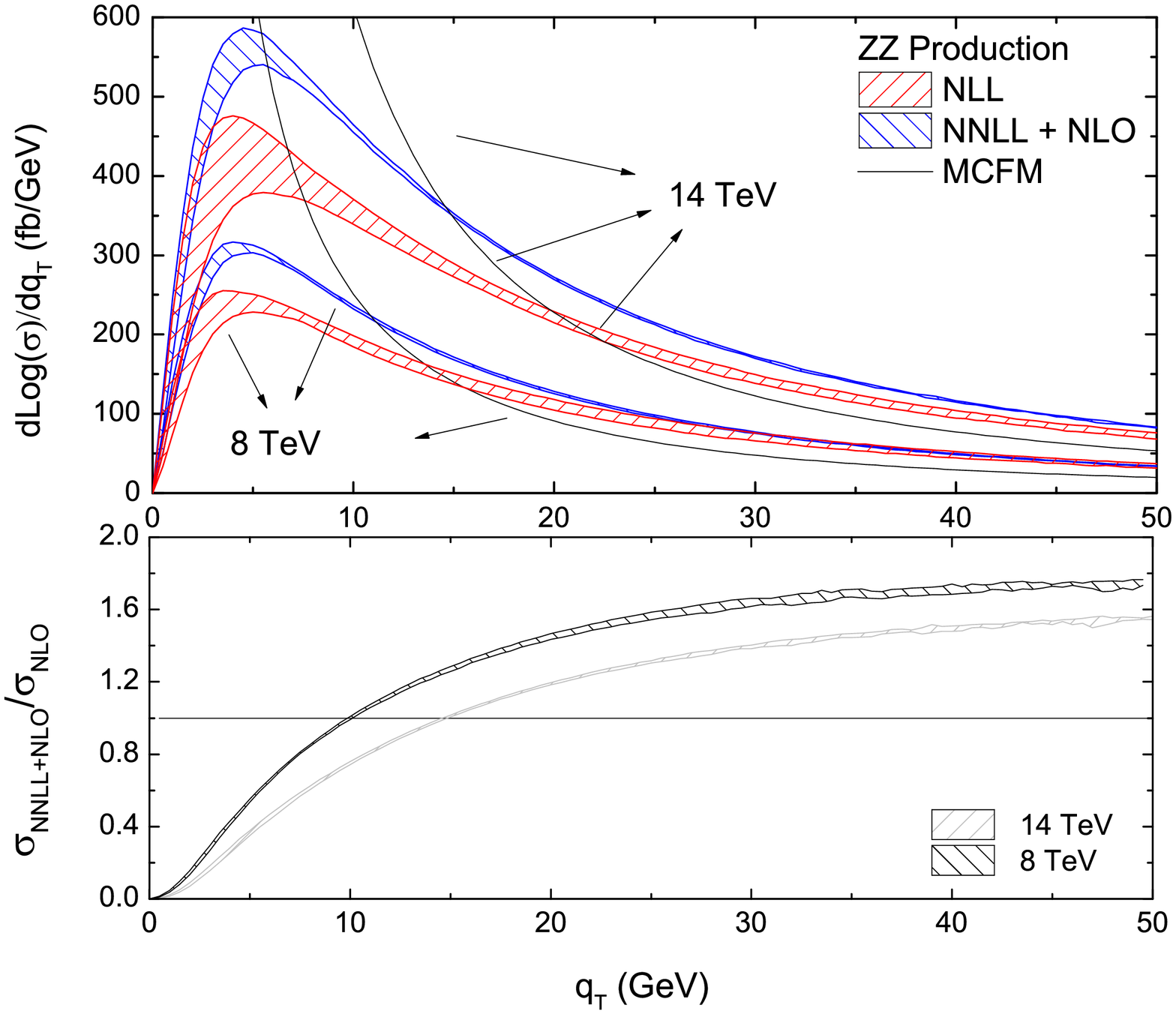}\\
\end{minipage}
\hfill
\begin{minipage}[t!]{0.45\linewidth}
\centering
 \includegraphics[width=1.0\linewidth]{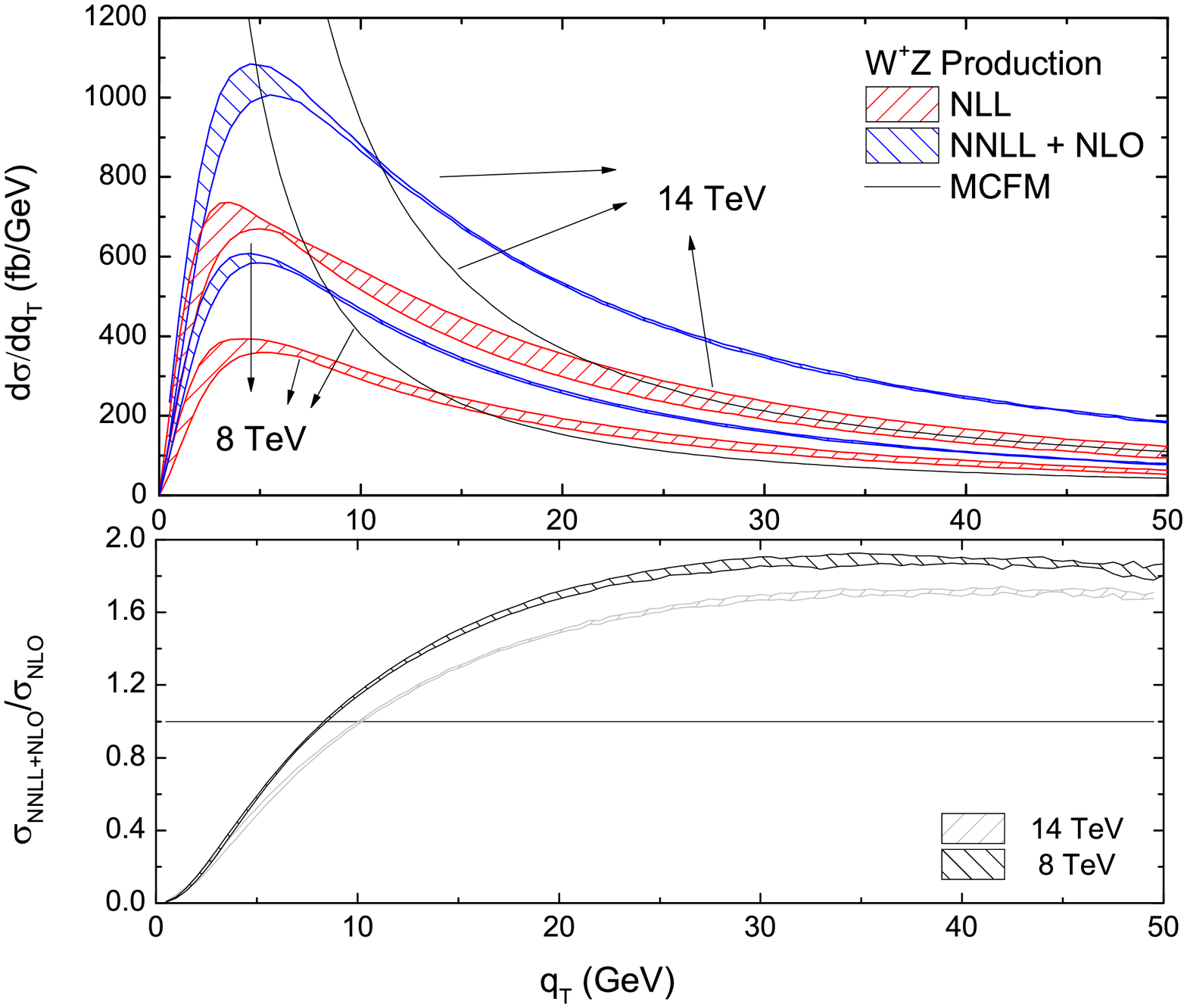}\\
\end{minipage}
\hfill
\begin{minipage}[t!]{0.45\linewidth}
\centering
 \includegraphics[width=1.0\linewidth]{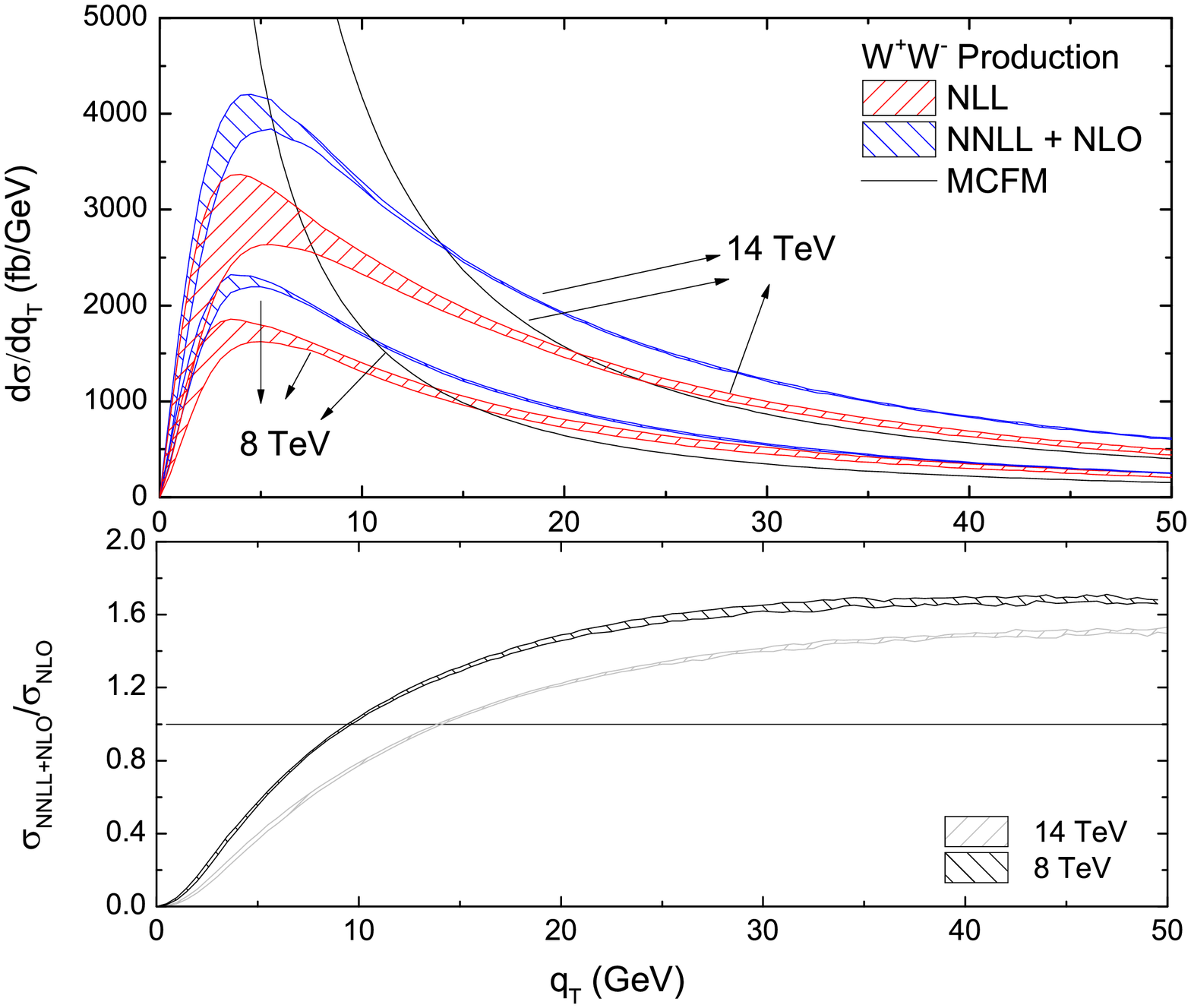}\\
\end{minipage}
\caption{Up: the $q_T$ distributions with scale uncertainties for gauge boson pair productions at the LHC with $\sqrt{S}=8~\text{TeV}$ and $\sqrt{S}=14~\text{TeV}$. Down: NNLL + NLO results normalized to NLO.}\label{f_fac_scale_dependence}
\end{figure}

Fig.~\ref{f_fac_scale_dependence} shows the resummed $q_T$ distributions for $W^+W^-,~ZZ$, and $W^{+}Z$ productions at next-to-leading logarithmic (NLL) and  NNLL + NLO accuracy at the LHC with $\sqrt{S}=8$ TeV and $\sqrt{S}=14$ TeV, respectively, which include the uncertainties of the theoretical predictions by varying the factorization scale $\mu_f$  by a factor of two around the default choice.  In these three cases of the gauge boson pair productions, the peak heights of the $q_T$ spectrums for $\sqrt{S}=14~\text{TeV}$ are much larger than that for $\sqrt{S}=8~\text{TeV}$, and the peak positions have a shift of about 0.5 GeV, respectively. Besides it is shown that, compared with the NLL results, the NNLL + NLO predictions significantly reduce the scale uncertainties, which make the theoretical predictions more reliable.
In Fig.~\ref{f_fac_scale_dependence} the K factors, defined as $\sigma_{\text{NNLL+NLO}}/\sigma_{\text{NLO}}$, are also shown. The fixed-order predictions is calculated by MCFM, and is invalid at small $q_T$ region.
In these three cases, at $q_T=50~\text{GeV}$, the K factors are  1.7-1.8 for $\sqrt{S}=8~\text{TeV}$ and  1.5-1.6 for $\sqrt{S}=14~\text{TeV}$, respectively.
And in the very large $q_T$ region resumed results agree with MCFM predictions. Note that, to our knowledge, the $q_T$ distribution of $W^{\pm}Z$ production at NNLL + NLO accuracy is not available  in both SCET and CSS frames before.

\begin{figure}[t!]
\begin{minipage}[t]{0.45\linewidth}
\centering
  \includegraphics[width=1.0\linewidth]{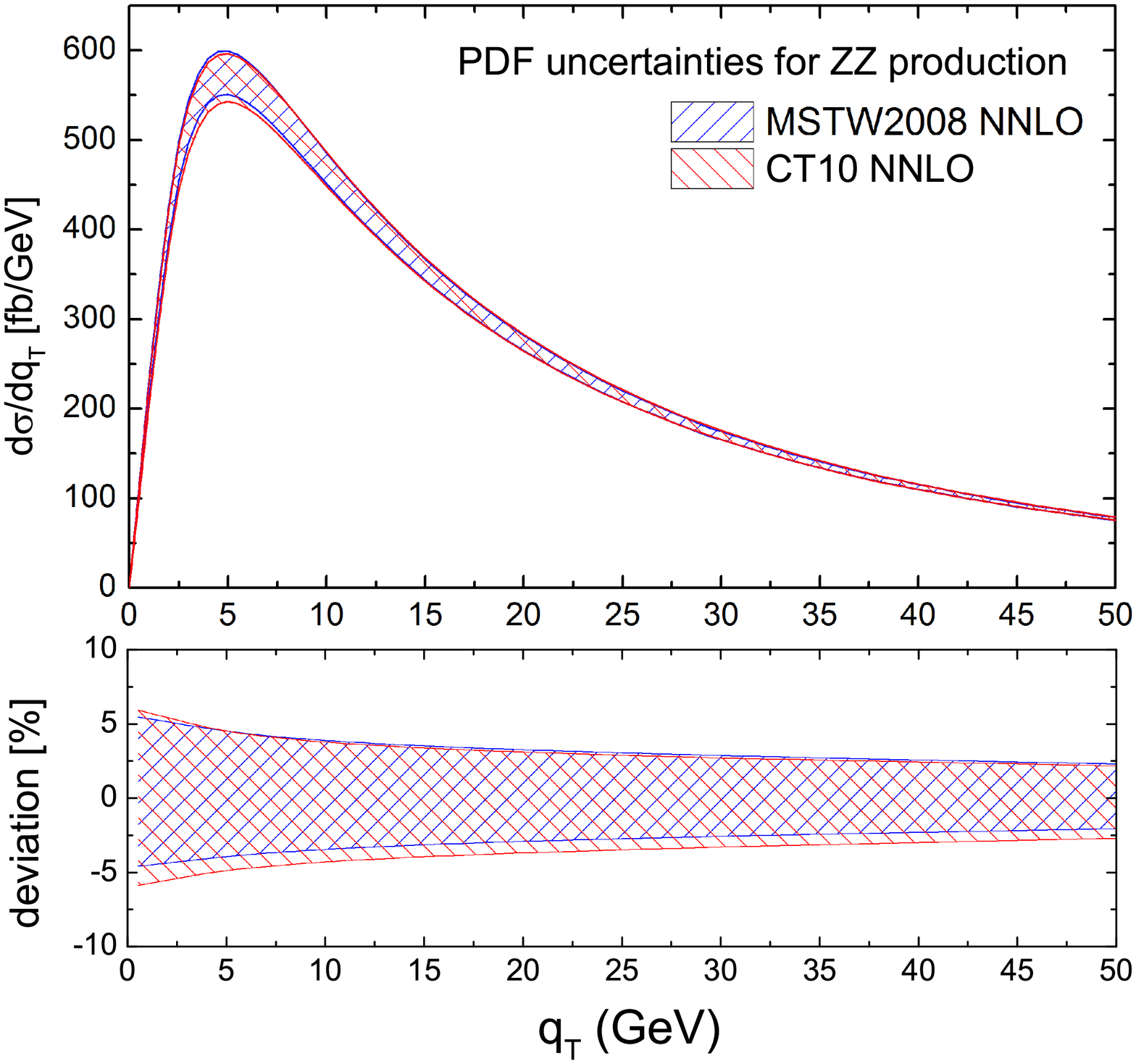}\\
\end{minipage}
\hfill
\begin{minipage}[t]{0.45\linewidth}
\centering
 \includegraphics[width=1.0\linewidth]{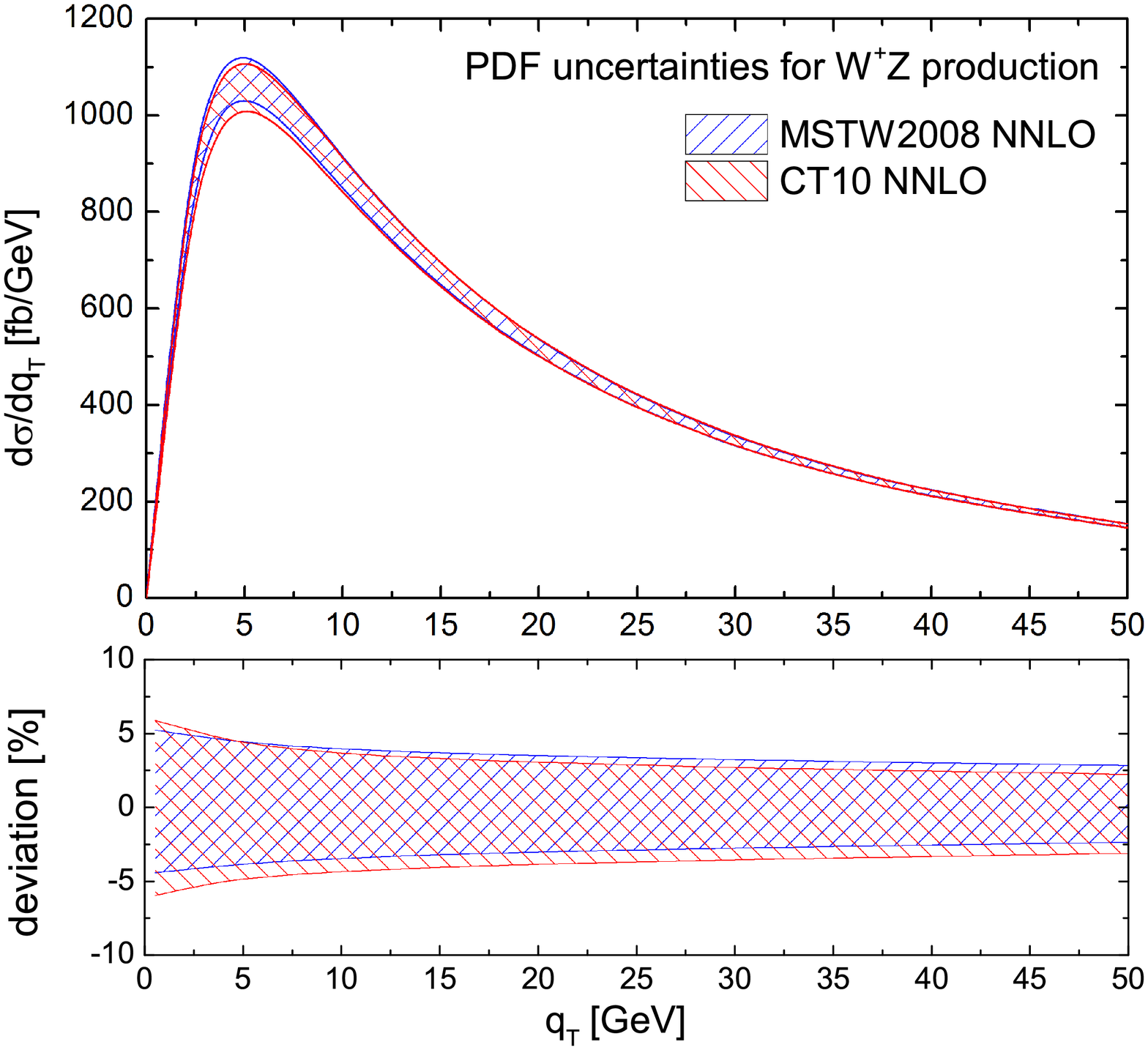}\\
\end{minipage}
\hfill
\begin{minipage}[t]{0.45\linewidth}
\centering
 \includegraphics[width=1.0\linewidth]{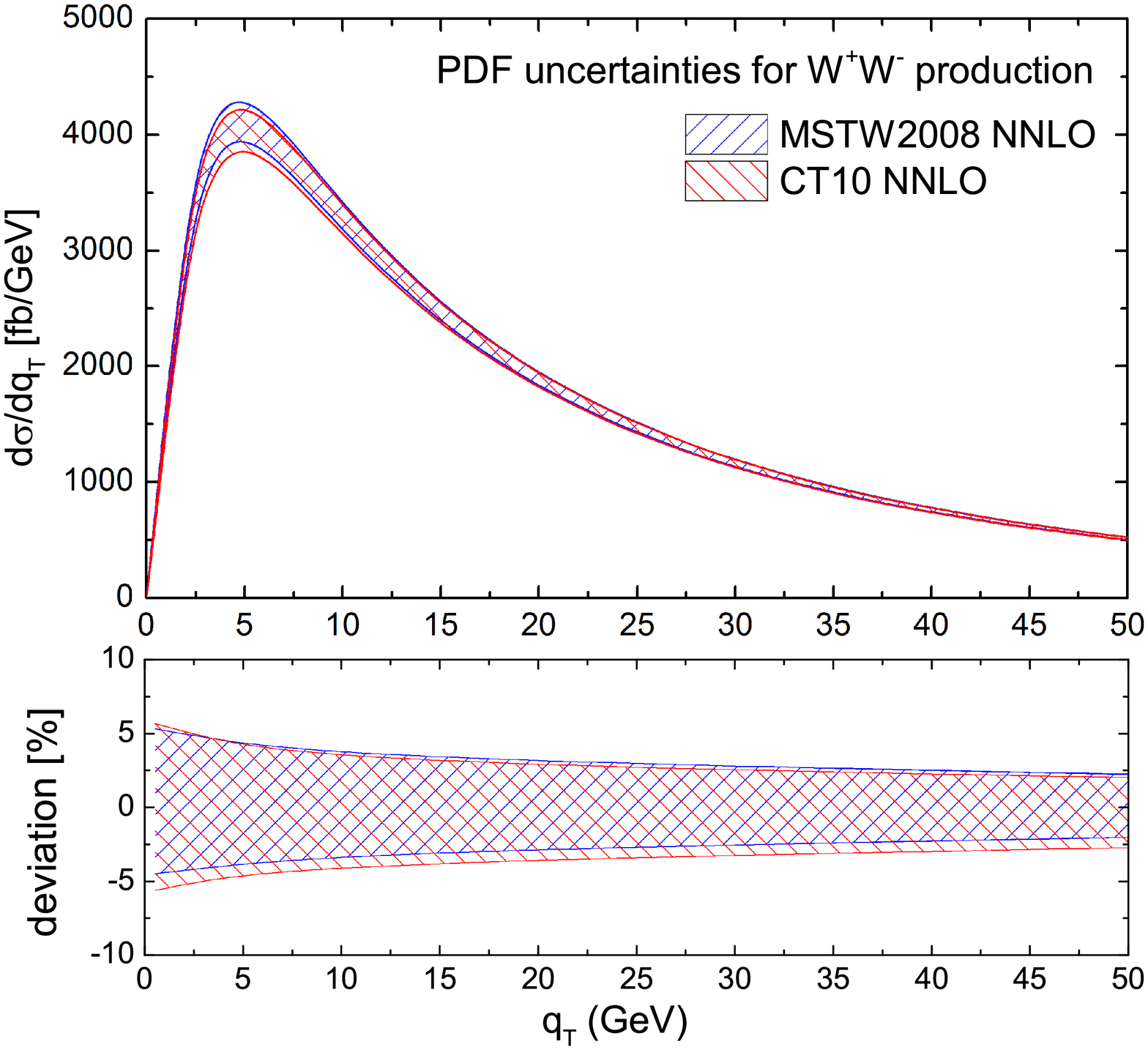}\\
\end{minipage}
\caption{\label{f_pdf}The PDF uncertainties of NNLL order resummed transverse-momentum for gauge boson pair production, where the bands represent 2$\sigma$ deviation.}
\end{figure}

In Fig.~\ref{f_pdf}, we show the PDF uncertainties of the NNLL order transverse-momentum distributions of the gauge boson pair at $2 \sigma$ deviation for $\sqrt{S}=14~\text{TeV}$.
The PDF uncertainties are of order 5\% at low $q_T$ region, and decrease to 2.5\% at high $q_T$ region for all three cases, respectively. We also show the PDF uncertainties with CT10NNLO PDF sets in the same plots, and the PDF uncertainties are a little larger than the cases with MSTW2008NNLO PDF sets. The situations for $\sqrt{S}=8~\text{TeV}$ are almost the same, and we do not discuss them here.

\begin{figure}[t!]
\begin{minipage}[t]{0.45\linewidth}
\centering
  \includegraphics[width=1.0\linewidth]{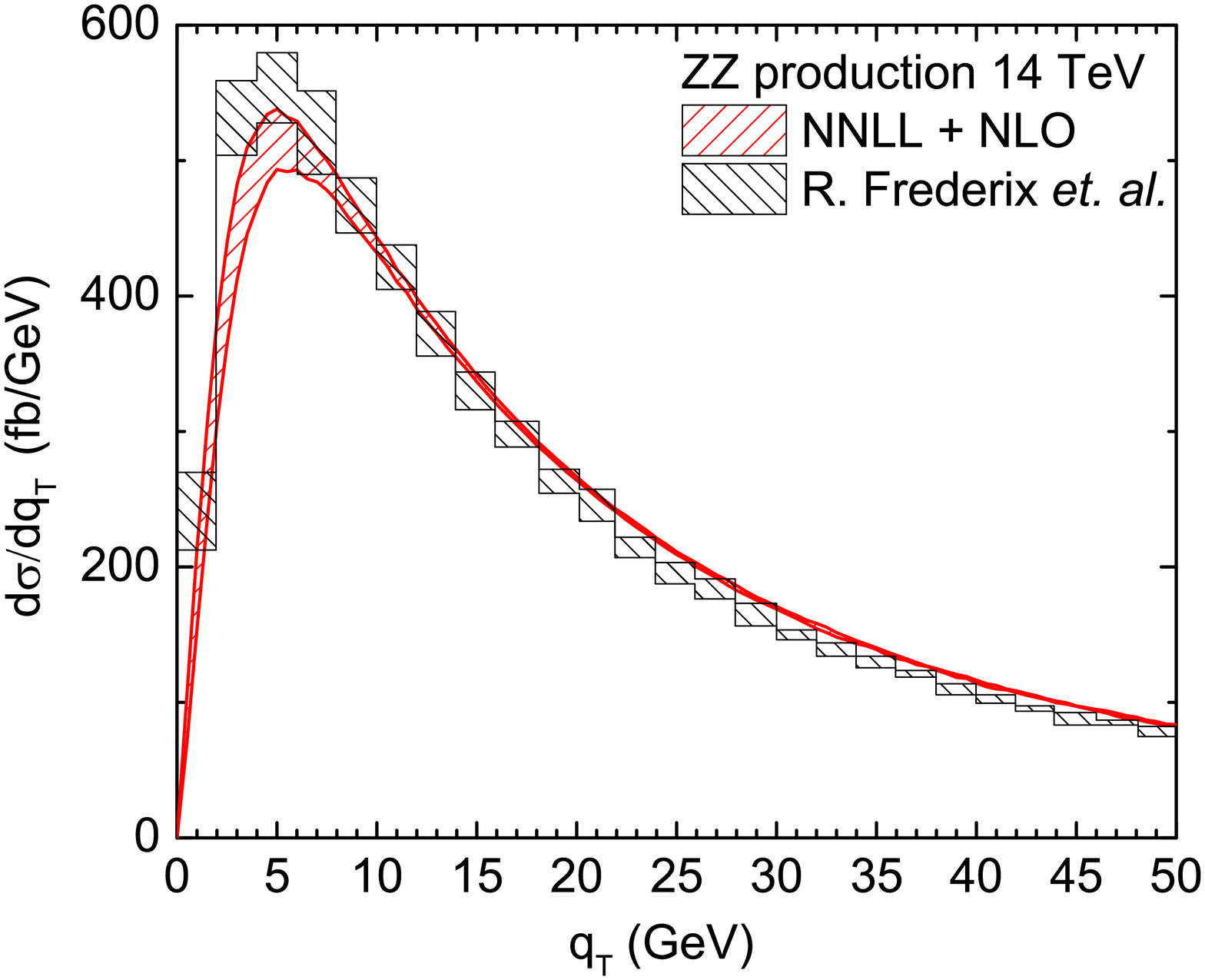}\\
\end{minipage}
\hfill
\begin{minipage}[t]{0.45\linewidth}
\centering
 \includegraphics[width=1.0\linewidth]{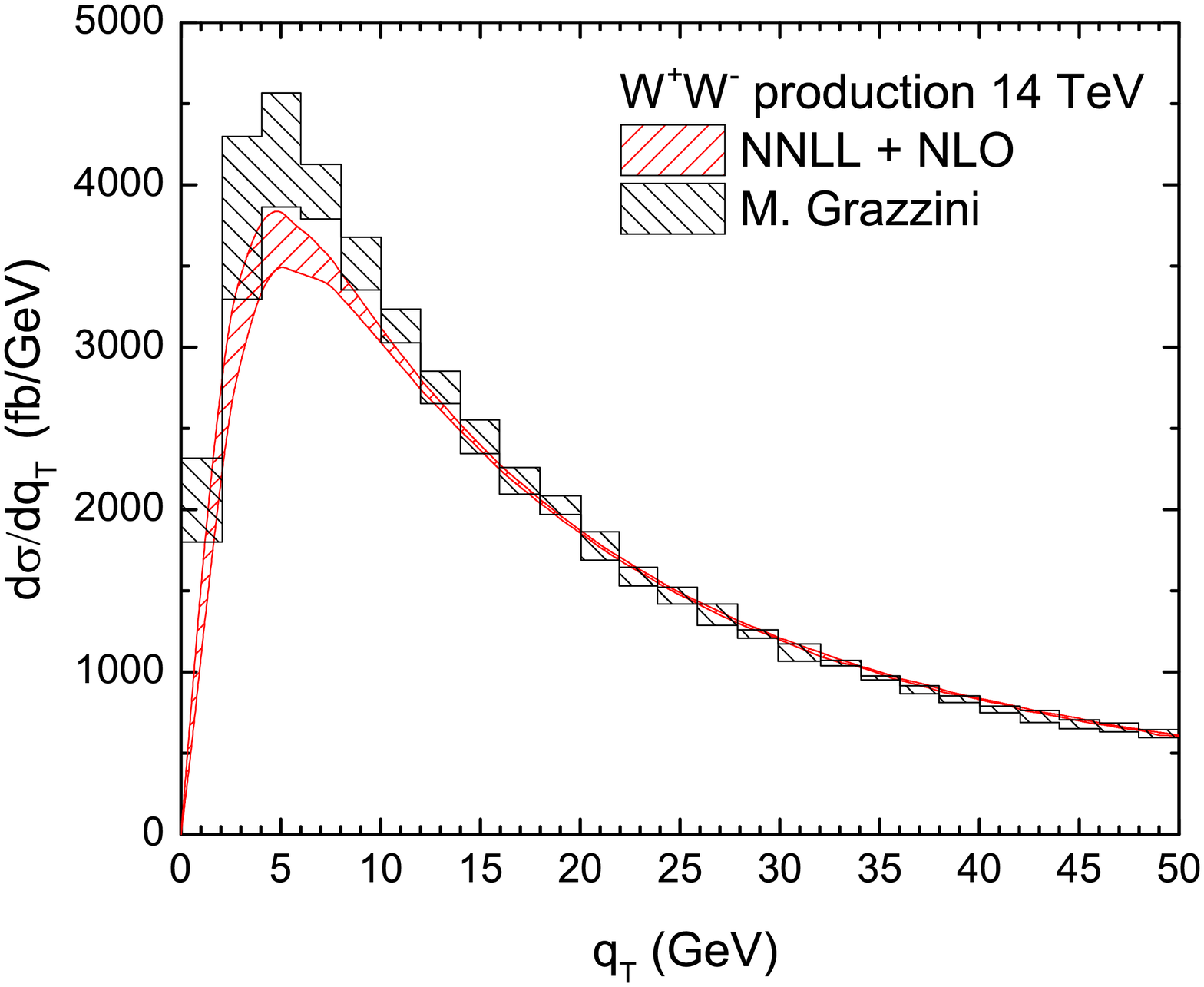}\\
\end{minipage}
\caption{Comparison of NNLL + NLO resummed $q_T$ distribution for $W^+W^-$ and $ZZ$  distributions in the SCET and CSS frame at $\sqrt{S}=14~\text{TeV}$ with MRST2002NLO PDF set. }\label{f_paper}
\end{figure}

In Fig.~\ref{f_paper}, we compare our NNLL + NLO results with previous studies~\cite{Grazzini2005a,Frederix2008} in CSS frame with MRST2002NLO PDF set for $\sqrt{S}=14~\text{TeV}$. The peak positions of the transverse-momentum spectrum for $W^+W^-$ and $ZZ$ productions in our results and those in CSS results are both at about 5 GeV. However, as shown in Fig.~\ref{f_paper}, in the small $q_T$ region, the peaks height in our results are  a little lower than those in CSS frame. Probably, this is due to the  fact that the choices of scales in two scheme are different. In CSS frame, the renormalization and the factorization scale are set to 2$m_{W/Z}$, and resummation scale is the invariant mass of the gauge boson pair. However in SCET frame there are only factorization scale and hard scale. The factorization scale is chosen as $q^* + q_T$, as described in Sec.~\ref{s2}, while the hard scale is taken as the invariant mass of the gauge boson pair. 

\begin{figure}[t!]
  \includegraphics[width=0.7\linewidth]{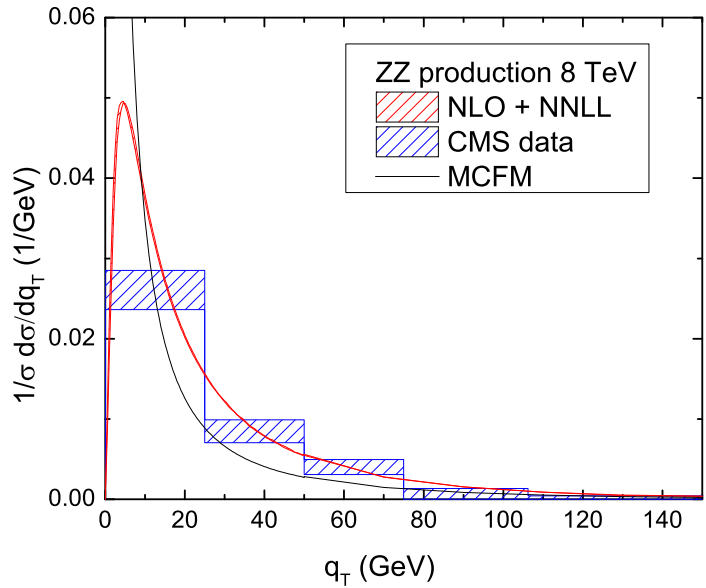}\\
  \caption{Comparison of normalized $q_T$ distribution for $ZZ$ productions between CMS experimental data and resummation prediction at the LHC with $\sqrt{S}=8$ TeV.}\label{f_exp}
\end{figure}

In Fig.~\ref{f_exp}, we compare the resummed results for the normalized differential cross section with the experimental data measured by the CMS collaboration~\cite{CMS-PAS-SMP-13-005} for $ZZ$ production  at the LHC with $\sqrt{S}=8$ TeV. Obviously, our NNLL + NLO predictions are consistent with the experimental data within theoretical and experimental uncertainties.
\section{Conclusion}\label{s4}
We have studied the  transverse-momentum resummation for $W^{+}W^{-}$, $ZZ$, and $W^{\pm}Z$ pair productions at the NNLL + NLO accuracy with SCET at the LHC. Especially, this is the first calculation of  $W^{\pm}Z$  transverse-momentum resummation at the NNLL + NLO accuracy. The non-perturbative effects are also included in our calculations. In these three cases of the gauge boson pair productions, our  results show that the peak positions  are all around $5~\text{GeV}$ for $\sqrt{S}=8~\text{TeV}$ and  $\sqrt{S}=14~\text{TeV}$, respectively, which agree quite well with previous results for $W^{+}W^{-}$ and  $ZZ$ productions, and the PDF uncertainties are less than 5\% at the $2\sigma$ level for the peak region. We also find that  our results agree well with experimental data reported by the CMS collaborations for the $ZZ$ productions at $\sqrt{S}=8~\text{TeV}$ within theoretical and experimental uncertainties.
\section{Acknowledgements}
This work was supported by the National Natural Science Foundation of China, under Grants No. 11021092 and No. 11135003.

\appendix
\section{Results of hard function}\label{appa}
Here, we show the detail results of $\mathcal{H}_{V_{l}V_{m}}(M,\mu_h)$ of the gauge boson pair production. We define
\begin{eqnarray}\label{s2_eq_HVV}
\mathcal{H}_{V_{l}V_{m}}(M,\mu_h)= \frac{1}{2s}\int H_{V_{l}V_{m}}(s,t,u,\mu_h) \frac{d^3 p_3}{(2\pi)^3 2E_3} \frac{d^3 p_4}{(2\pi)^3 2E_4} (2\pi)^4\delta^4(q-p_3-p_4),
\end{eqnarray}
and expand $H_{V_{l}V_{m}}$ as
\begin{eqnarray}
H_{V_{l}V_{m}} = H_{V_{l}V_{m}}^{(0)} + \frac{\alpha_s}{2\pi}H_{V_{l}V_{m}}^{(1)}.
\end{eqnarray}
The leading order coefficient is
\begin{eqnarray}
H_{V_{l}V_{m}}^{(0)} = \overline{|\mathcal{M}^B_{V_{l}V_{m}}|^2},
\end{eqnarray}
where $\overline{|\mathcal{M}^B_{V_{l}V_{m}}|^2}$ is the color-averaged and spin-averaged tree-level matrix element squared.
$H_{{V_{l}V_{m}}}^{(1)}$ can be divided into two parts
\begin{eqnarray}
H_{V_{l}V_{m}}^{(1)} = \frac{\alpha_s}{2\pi} \left(H_{V_{l}V_{m},\text{reg}}^{(1)} +H_{\mu}^{(1)}\right).
\end{eqnarray}
where $H_{\mu}^{(1)}$ has the same form for three cases:
\begin{eqnarray}
H_{\mu}^{(1)} &=& C_F H_{V_{l}V_{m}}^{(0)} \left(-\ln^2
\frac{M^2}{\mu_h^2}+3\ln\frac{M^2}{\mu_h^2}-\frac{\pi^2}{6}\right).
\end{eqnarray}
For simplicity, we define that all scalar one-loop integrals should be understood as retaining the finite part, and
\begin{eqnarray}
\text{C}_{0}^{s}&=&\text{C}_0\left(0,0,s,0,0,0\right),\nonumber\\
\text{C}_0^{V_1,V_2}&=&\text{C}_0\left(m_{V_1}^2,s,m_{V_2}^2,0,0,0\right),\nonumber\\
\text{C}^{V,t}_0&=&\text{C}_0\left(0,m_V^2,t,0,0,0\right),\nonumber\\
\text{D}_0^{V_1,V_2}&=&\text{D}_0\left(0,0,m_{V_1}^2,m_{V_2}^2,s,t,0,0,0,0\right).
\end{eqnarray}
\subsection{$ZZ$ production}
The leading order (LO) coefficient is~\cite{Mele1991409}
\begin{eqnarray}
H_{ZZ}^{(0)} &=&\frac{1}{N_c} ((g_{i,Z}^{L})^4+(g_{i,Z}^{R})^4) \times \nonumber\\
&& \left(\frac{-\left(m_Z\right){}^4 \left(t^2-8 t
   u+u^2\right)-4 t u \left(m_Z\right){}^2 (t+u)+t u
   \left(t^2+u^2\right)}{t^2 u^2}\right).
\end{eqnarray}
where $g_{i,Z}^{L,R},~i=u,d$ is the couplings between the $Z$ boson and quarks
\begin{eqnarray}
g_{u,Z}^{L}&=&\frac{e}{\sin\theta_{W}\cos\theta_{W}}\left(\frac{1}{2}-\frac{2\sin^2\theta_{W}}{3}\right),\nonumber\\
g_{d,Z}^{L}&=&\frac{e}{\sin\theta_{W}\cos\theta_{W}}\left(-\frac{1}{2}+\frac{\sin^2\theta_{W}}{3}\right),\nonumber\\
g_{u,Z}^{R}&=&-\frac{2}{3}\tan\theta_{W},\nonumber\\
g_{d,Z}^{R}&=&-\frac{1}{3}\tan\theta_{W},
\end{eqnarray}
where $\theta_{W}$ is the Weinberg angle. The functions at the $\mathcal{O}(\alpha_s)$ are~\cite{Mele1991409},
\begin{eqnarray}
\frac{\alpha_s}{2\pi} H_{ZZ,\text{reg}}^{(1)}&=&\frac{1}{12}\frac{\alpha _s }{2\pi}C_F((g_{i,Z}^{L})^4+(g_{i,Z}^{R})^4)\nonumber\\
&&\bigg(\text{A1}(t,u)+\text{A2}(t,u) \ln\left(-\frac{m_Z^2}{t}\right) +
         \text{A3}(t,u)\ln\left(-\frac{m_Z^2}{s}\right)+\text{A4}(t,u)\text{D}_0^{Z,Z}\nonumber\\
         && +\text{A5}(t,u) \text{C}^{Z,t}_0+\text{A6}(t,u)\text{C}_0^{Z,Z}+\text{A7}(t,u)\text{C}_{0}^{s} +(t\Leftrightarrow u)\bigg),
\end{eqnarray}
where
\begin{eqnarray}
A1(t,u) &=& \frac{2 \left(m_{Z}^2+t\right)^2}{s t u \left(1-\frac{4 m_{Z}^2}{s}\right)}+\frac{4 s}{u \left(t-m_{Z}^2\right)}\nonumber\\
&~&-\frac{-36 m_{Z}^6+18 m_{Z}^4 s+t^2 \left(28 s-68 m_{Z}^2\right)+t \left(88 m_{Z}^4-36 m_{Z}^2 s+18 s^2\right)+16 t^3}{s t^2 u},\nonumber\\
\end{eqnarray}
\begin{eqnarray}
A2(t,u) = -\frac{6 \left(m_{Z}^4+s^2\right)}{s t u}-\frac{4 m_{Z}^2 s}{u \left(t-m_{Z}^2\right)^2}+\frac{12 s}{u \left(t-m_{Z}^2\right)}+\frac{6 m_{Z}^4
   \left(2 m_{Z}^2-s\right)}{s t^2 u}+\frac{2 (4 s+t)}{s u},\nonumber\\
\end{eqnarray}
\begin{eqnarray}
A3(t,u) &=& -\frac{-12 m_{Z}^6+t \left(25 m_{Z}^4+6 s^2\right)+6 m_{Z}^4 s+t^2 \left(8 s-18 m_{Z}^2\right)+5 t^3}{s t^2 u}\nonumber\\
&~&\frac{-25 m_{Z}^4-26 m_{Z}^2 t+3 t^2}{s t u \left(1-\frac{4 m_{Z}^2}{s}\right)}-\frac{12 m_{Z}^2 \left(m_{Z}^2+t\right)^2}{s^2 t u
   \left(1-\frac{4 m_{Z}^2}{s}\right)^2},
\end{eqnarray}
\begin{eqnarray}
A4(t,u) &=& \frac{-12 m_{Z}^4+8 m_{Z}^2 t-2 s^2-2 t^2}{u}+\frac{4 m_{Z}^4 \left(2 m_{Z}^2-s\right)}{t u},
\end{eqnarray}
\begin{eqnarray}
A5(t,u) &=& \frac{8 m_{Z}^6 \left(2 m_{Z}^2-s\right)}{s t^2 u}+\frac{4 \left(10 m_{Z}^4-5 m_{Z}^2 t+s^2+t^2\right)}{s u}\nonumber\\
&&+\frac{4 m_{Z}^2 \left(-10 m_{Z}^4+2 m_{Z}^2
   s-s^2\right)}{s t u},
\end{eqnarray}
\begin{eqnarray}
A6(t,u) &=& \frac{-3 m_{Z}^6+12 m_{Z}^4 s-4 m_{Z}^2 s^2+t^2 \left(2 s-3 m_{Z}^2\right)+t \left(6 m_{Z}^4-8 m_{Z}^2 s\right)+2 s^3}{s t u}\nonumber\\
&~& +\frac{12 m_{Z}^4 \left(m_{Z}^2+t\right)^2}{s^2 t u \left(1-\frac{4 m_{Z}^2}{s}\right)^2}-\frac{-27 m_{Z}^6-30 m_{Z}^4 t+m_{Z}^2 t^2}{s t u
   \left(1-\frac{4 m_{Z}^2}{s}\right)},
\end{eqnarray}
\begin{eqnarray}
A7(t,u) &=& \frac{4}{3} \left(\frac{\left(s-2 m_{Z}^2\right)^2}{s t u}+\frac{-4 m_{Z}^2+4 s+t}{s u}\right).
\end{eqnarray}
\subsection{$W^{\pm}Z$ production}
The LO results can be expressed as~\cite{Frixione:1992pj}:
\begin{eqnarray}
H_{WZ}^{(0)}&=&\frac{1}{6} \left(\frac{1}{2\sqrt{2}\sin\theta_{W}}\right)^2 \Bigg((g_{d,Z}^L)^2 I_{dd}^{(0)}(s,t,u) + 2 g_{d,Z}^L g_{u,Z}^L I_{ud}^{(0)}(s,t,u) + (g_{u,Z}^L)^2 I_{dd}^{(0)}(s,u,t) \nonumber\\
&~&+ 2 g_{W,Z}g_{d,Z}(F_{d}^{(0)}(s,t,u)-F_{d}^{(0)}(s,u,t)) + g_{W,Z}^2J^{(0)}(s,t,u)\Bigg),
\end{eqnarray}
where
\begin{eqnarray}
g_{W,Z}&=&\frac{-e}{s-m_W^2}\frac{\cos\theta_W}{\sin\theta_W},
\end{eqnarray}
and
\begin{eqnarray}
I_{dd}^{(0)}(s,t,u) &=& 8 \left(\frac{s \left(m_{W}^2+m_{Z}^2\right)}{2 m_{W}^2 m_{Z}^2}+\frac{1}{4} \left(\frac{t u}{m_{W}^2 m_{Z}^2}-1\right)\right)+8
   \left(\frac{u}{t}-\frac{m_{W}^2 m_{Z}^2}{t^2}\right),
\end{eqnarray}
\begin{eqnarray}
I_{ud}^{(0)}(s,t,u) &=& \frac{8 s \left(m_{W}^2+m_{Z}^2\right)}{t u}-8 \left(\frac{s \left(m_{W}^2+m_{Z}^2\right)}{2 m_{W}^2 m_{Z}^2}+\frac{1}{4} \left(\frac{t u}{m_{W}^2
   m_{Z}^2}-1\right)\right),
\end{eqnarray}
\begin{eqnarray}
F_{d}^{(0)}(s,t,u) &=&-8 s \Bigg(\frac{1}{4} \left(-\frac{4 m_{W}^2 m_{Z}^2}{s t}-\frac{m_{W}^2+m_{Z}^2}{s}+1\right) \left(\frac{t u}{m_{W}^2 m_{Z}^2}-1\right)\nonumber\\
&~& +\frac{\left(m_{W}^2+m_{Z}^2\right) \left(\frac{2 m_{W}^2 m_{Z}^2}{t}-m_{W}^2-m_{Z}^2+s\right)}{2 m_{W}^2 m_{Z}^2}\Bigg),
\end{eqnarray}
\begin{eqnarray}
J^{(0)}(s,t,u) &=&8 s^2 \left(\frac{8 m_{W}^2 m_{Z}^2+\left(m_{W}^2+m_{Z}^2\right)^2}{4 s^2}-\frac{m_{W}^2+m_{Z}^2}{2 s}+\frac{1}{4}\right) \left(\frac{t u}{m_{W}^2
   m_{Z}^2}-1\right) \nonumber\\
   &~& +\frac{8 s^2 \left(m_{W}^2+m_{Z}^2\right) \left(\frac{\left(m_{W}^2-m_{Z}^2\right)^2}{2 s}-m_{W}^2-m_{Z}^2+\frac{s}{2}\right)}{m_{W}^2 m_{Z}^2}.
\end{eqnarray}

When considering virtual corrections, as in the tree level case, we have~\cite{Frixione:1992pj}
\begin{eqnarray}
\frac{\alpha_s}{2\pi} H_{WZ,\text{reg}}^{(1)} &=& \frac{1}{6}\frac{\alpha _s }{2\pi}C_F\left(\frac{1}{2\sqrt{2}\sin\theta_{W}}\right)^2 \nonumber\\&&
\Bigg((g_{d,Z}^L)^2 I_{dd}^{(1)}(s,t,u) + g_{d,Z}^Lg_{u,Z}^L I_{ud}^{(1)}(s,t,u) + (g_{u,Z}^L)^2 I_{dd}^{(1)}(s,u,t) \nonumber\\
&~&+ g_{W,Z}g_{d,Z}(F_{d}^{(1)}(s,t,u)-F_{d}^{(1)}(s,u,t)) + g_{W,Z}^2J^{(1)}(s,t,u)\Bigg),
\end{eqnarray}
where
\begin{eqnarray}
I_{dd}^{(1)} &=& \frac{2 \left(22 t^2+t (19 s-18 \Sigma )+ 18 m_W^2
   m_Z^2\right)}{t^2} -\frac{8 (u t+2 s \Sigma )}{m_W^2 m_Z^2}-\frac{2 (t-u)^2}{t s \beta ^2}\nonumber\\
   &&+\Bigg(\frac{2
   \left(8 t^2+4 t (s-3 \Sigma )+4 \Sigma ^2-5 s \Sigma +s^2\right)}{t s \beta ^2}+\frac{4 \left(t (3 u+s)-3 m_W^2
   m_Z^2\right)}{t^2}\nonumber\\
   &&+\frac{6 (t+u) (t-u)^2}{t s^2 \beta ^4}\Bigg) \ln \left(-\frac{t}{s}\right)\nonumber\\
   &&+\Bigg(\frac{8 t^2
   (-2 s+\Delta )+8 t \left(-s^2+3 s \Sigma -2 \Delta  \Sigma \right)-2 (s-\Sigma ) \left(s^2-4 s \Sigma +3 \Delta  \Sigma
   \right)}{t s^2 \beta ^2}\nonumber\\
   &&+\frac{16 s \left(t-m_Z^2\right)}{t (u+s)-m_W^2 m_Z^2}-\frac{6 (s-\Delta )(t+u) (t-u)^2}{t s^3 \beta^4}\nonumber\\
   &&+\frac{2 \left(4 t^2+t
   \left(10 s-3 m_Z^2-9 m_W^2\right)+12 m_W^2 m_Z^2\right)}{t^2}\Bigg) \ln \left(-\frac{t}{m_W^2}\right)\nonumber\\
   &&+\bigg(-\frac{4 t^2 (2 \Sigma -3 s)-4 t (s-\Sigma ) (2 s-3
   \Sigma )-2 (s-2 \Sigma ) (s-\Sigma )^2}{t s \beta ^2}\nonumber\\
   &&+\frac{4 \Sigma  t-3 s^2+4 s \Sigma -4
   \left(m_W^4+m_Z^4\right)}{t}-\frac{3 \left(t^2-u^2\right)^2}{t s^2 \beta ^4}\bigg)
   \text{C}_0^{W,Z}\nonumber\\
   &&+\left(\frac{4 (u t+2 s \Sigma )}{3 m_W^2 m_Z^2}-\frac{4(t-2 u)}{3 t}\right) \text{C}_{0}^{s}-\frac{4 s \left(t u-2 m_W^2 m_Z^2\right)
   \text{D}_0^{W,Z}}{t}\nonumber\\
   &&+\frac{\left(8 \left(t-m_W^2\right) \left(u t-2
   m_W^2 m_Z^2\right)\right) \text{C}^{W,t}_0}{t^2} + (m_W\Leftrightarrow m_Z),
\end{eqnarray}
\begin{eqnarray}
I_{ud}^{(1)} &=& \frac{4 s^2 (2 t-\Sigma )}{u \left(m_W^2 m_Z^2-t (s+u)\right)}+\frac{8 (2 s \Sigma +t u)}{m_W^2
   m_Z^2}+\frac{2 \left(-18 s \Sigma +t (9 s-4 \Sigma )+4 t^2\right)}{t u}\nonumber\\
   && +\Bigg(\frac{2 \left(-(s-\Sigma ) (3 s+4 \Sigma )-4 t
   (s+3 \Sigma )+8 t^2\right)}{\beta ^2 s u}\nonumber\\
   &&+\frac{6 (t+u) (t-u)^2}{\beta ^4 s^2 u}-\frac{12 s (t-\Sigma )}{t u}\Bigg)\ln \left(-\frac{t}{s}\right) \nonumber\\
   &&\Bigg(\frac{2}{u s^2 \beta^2} \Big(4 t^2 (-2 s + \Delta) +
   4 t (s^2 + s ( m_{Z}^2 + 5 m_{W}^2) -
      2 \Delta \Sigma) \nonumber\\
   &&+ (s - \Sigma) (3 \
s^2 + 8 s m_{W}^2 - 3 \Delta \Sigma)\Big)-\frac{8 s^2 t \left(t-m_Z^2\right) (2 t-\Sigma )}{u \left(m_W^2 m_Z^2-t (s+u)\right)^2}\nonumber\\
 &&+\frac{2 t \left(2
   m_W^2+m_Z^2+18 s\right)-24 s \Sigma }{t u}+\frac{6 (s-\Delta ) (s-\Sigma ) (t-u)^2}{\beta ^4 s^3 u}\nonumber\\
   &&-\frac{8 s \left(-t \left(2 m_W^2+4 m_Z^2+3 s\right)+2 m_Z^2 (s+\Sigma )+2 t^2\right)}{u \left(m_W^2
   m_Z^2-t (s+u)\right)}\Bigg)\ln \left(-\frac{t}{m_W^2}\right)\nonumber\\
   &&\Bigg(-\frac{2 \left((s-\Sigma )^2 (s+2 \Sigma )+2 t^2 (2 \Sigma -3 s)+6 \Sigma  t
   (s-\Sigma )\right)}{\beta ^2 s u}\nonumber\\
   &&+\frac{3 s (-s-4 \Sigma +4 t)}{u}-\frac{2 (t-u)^2}{\beta ^2 s u}-\frac{3 (s-\Sigma )^2 (t-u)^2}{\beta ^4 s^2 u}\Bigg)\text{C}_0^{W,Z}+\frac{8 s^2 (t-\Sigma )}{u} \text{D}_0^{W,Z}  \nonumber\\
   && +\left(\frac{4 (4 s+u)}{3 u} -\frac{4 (2 s \Sigma +t u)}{3 m_W^2 m_Z^2}\right)\text{C}_{0}^{s}+\frac{16 s \left(t-m_W^2\right) (t-\Sigma )}{t u}\text{C}^{W,t}_0\nonumber\\
   &&+ (t\Leftrightarrow u) +  (m_W\Leftrightarrow m_Z)+ (t\Leftrightarrow u,m_W\Leftrightarrow m_Z),
\end{eqnarray}
\begin{eqnarray}
F_{d}^{(1)} &=&\frac{4 \left(17 \left(m_{W}^2 m_{Z}^2+s \Sigma \right)+t (11 s-13 \Sigma )+17 t^2\right)}{t}\nonumber\\
&& +\frac{16
   (s-\Sigma ) (2 s \Sigma +t u)}{m_{W}^2 m_{Z}^2}+\frac{4 s^2 (2 t-\Sigma )}{t (s+u)-m_{W}^2 m_{Z}^2}\nonumber\\
   &&+ \left(\frac{8 (t-u)}{\beta ^2}-\frac{4 \left(3 \left(m_{W}^2 m_{Z}^2+s \Sigma \right)-t (s+3 \Sigma )+3 t^2\right)}{t^2}\right)\ln \left(-\frac{t}{s}\right)\nonumber\\
   &&\Bigg( \frac{8 \left(3 \left(m_{W}^2 m_{Z}^2+s \Sigma \right)-t \left(3 m_{W}^2+m_{Z}^2+2 s\right)+t^2\right)}{t}+\frac{8 s \left(t (3 s+2 \Sigma )-2 m_{Z}^2
   (s+\Sigma )\right)}{t (s+u)-m_{W}^2 m_{Z}^2}\nonumber\\
   &&\frac{8 s^2 t \left(t-m_{Z}^2\right) (2 t-\Sigma )}{\left(t (s+u)-m_{W}^2 m_{Z}^2\right)^2}-\frac{8 (s-\Delta ) (t-u)}{\beta ^2 s}\Bigg)\ln \left(-\frac{t}{m_W^2}\right)\nonumber\\
   &&4 \left(-m_{W}^4-m_{Z}^4+4 s \Sigma \right)+4 t (\Sigma -3 s)+\frac{4 (s-\Sigma ) (t-u)}{\beta ^2}\text{C}_0^{W,Z} \nonumber\\
   &&-\frac{8 \left(2 \left(m_{W}^2 m_{Z}^2+s \Sigma \right)+2 t (2 s-\Sigma )+3 t^2\right)}{3 t}+\frac{8 (s-\Sigma ) (2 s \Sigma +t u)}{3 m_{W}^2 m_{Z}^2}\text{C}_{0}^{s} \nonumber\\
   &&4 \left(2 s \left(m_{W}^2 m_{Z}^2+s \Sigma \right)+s t^2-s t (s+\Sigma )\right) \text{D}_0^{W,Z}\nonumber\\
   &&-\frac{8 \left(t-m_{W}^2\right) \left(2 \left(m_{W}^2 m_{Z}^2+s \Sigma \right)-t (s+\Sigma )+t^2\right)}{t} \text{C}^{W,t}_0 + (m_W\Leftrightarrow m_Z),
\end{eqnarray}
\begin{eqnarray}
 J^{(1)} &=& \left(16-\frac{8 \pi ^2}{3}\right) \big(m_W^4-\frac{(s-\Sigma )^2 (2 s \Sigma +t u)}{m_W^2
   m_Z^2} \nonumber\\
   && +10 m_W^2 m_Z^2+m_Z^4+s^2+6 s \Sigma +8 t (s-\Sigma )+8 t^2\big),
\end{eqnarray}
with
\begin{eqnarray}
\Sigma &=& m_Z^2+m_W^2,\nonumber\\
\Delta &=& m_Z^2-m_W^2,\nonumber\\
\beta &=& \sqrt{1-\frac{(m_W+m_Z)^2}{s}}\sqrt{1-\frac{(m_W-m_Z)^2}{s}}.
\end{eqnarray}

\subsection{$W^+W^-$ production}
The LO results are~\cite{Frixione:1993yp}
\begin{eqnarray}
H_{WW}^{(0)}=\frac{1}{12}\left(c_q^{tt}F_q^0(s,t)+c_q^{ss}K_q^0(s,t)-c_q^{ts}J_q^0(s,t)\right).
\end{eqnarray}
The coefficients  are
\begin{eqnarray}
c_q^{tt}&=&{\pi^2\alpha^2\over \sin^2\theta_W},\nonumber \\
c_q^{ts}(s)&=&{4\pi^2\alpha^2\over  \sin^2\theta_W}
{1\over s}
\biggl(Q_q+{s\over s-m_Z^2}
{1\over \sin^2\theta_W}(T_{3,q}-Q_q \sin^2\theta_W)\biggr),\nonumber \\
c_q^{ss}(s)&=&{16\pi^2\alpha^2\over s^2}\biggl\{
\biggl(Q_q+{1\over 2 \sin^2\theta_W}(T_{3,q}-2Q_q\sin^2\theta_W){s\over s-m_Z^2}\biggr)^2
\nonumber\\
&&+\biggl({T_{3,q}\over 2 \sin^2\theta_W}{s\over s-m_Z^2}\biggr)^2\biggr\},
\end{eqnarray}
with $T_{3,q}=\pm {1\over 2}$.
The functions occurring in the lowest order amplitudes are,
\begin{eqnarray}
F_u^{0}(s,t)&=&F_d^{0}(s,u)
\nonumber \\
&=&16\biggl(
{ut\over m_W^4}-1
\biggr)
\biggl({1\over 4}+{m_W^4\over t^2}
\biggr)+16{s\over m_W^2},
\nonumber \\
J_u^{0}(s,t)&=&-J_d^{0}(s,u)
\nonumber \\ &=&
16\biggl(
{ut\over m_W^4}-1\biggr)\biggl({s\over 4}-{m_W^2\over 2}-{m_W^4\over t}\biggr)
+16s\biggl({s\over m_W^2}-2+{2m_W^2\over t}\biggr),
\nonumber \\
K_u^{0}(s,t)&=& K_d^{0}(s,u)\nonumber \\
&=&
8\biggl({ut\over m_W^4}-1\biggr)
\biggl({s^2\over 4}-sm_W^2+3m_W^4\biggr)
+8s^2\biggl({s\over m_W^2}-4\biggr)
\, .
\end{eqnarray}
The functions at the $\mathcal{O}(\alpha_s)$ are~\cite{Frixione:1993yp},
\begin{eqnarray}
\frac{\alpha_s}{2\pi} H_{WW,\text{reg}}^{(1)}= \frac{1}{24}\frac{\alpha _s }{2\pi}C_F\left(c_q^{tt}F_q^1(s,t)+c_q^{ss}K_q^1(s,t)-c_q^{ts}J_q^1(s,t)\right),
\end{eqnarray}
where
\begin{eqnarray}
F_u^{1}(s,t)&=&{4(80t^2+73st-140m_W^2t+72m_W^4 \over t^2}
-{4(4t+s)^2\over s\beta^2 t}
-{128(t+2s)\over m_W^2}\nonumber \\&&
+{64(t+s)\over m_W^4}
-\biggl({32(t^2-3st-3m_W^4)\over t^2}+{128s\over t-m_W^2}\biggr)\ln\biggl({-t\over m_W^2}\biggr)
\nonumber \\
&&+\biggl({8(6t^2+8st-19m_W^2t+12m_W^4)\over t^2}-{32t^2-128st-26s^2\over s\beta^2t}
\nonumber \\
&&+{6(4t+s)^2\over s\beta^4t}\biggr)\ln\biggl({s\over m_W^2}\biggr)
+32s\biggl({2m_W^4\over t}-u\biggr)\text{D}_0^{W,W}\nonumber \\
&&
-64 (t-m_W^2)\biggl(
{2m_W^4\over t^2}-{u\over t}\biggr)\text{C}^{W,t}_0
\nonumber \\
&&+\biggl(
{16t(4m_W^2-u)-49s^2+72m_W^2s-48m_W^4\over 2t}
+{2(8t^2-14st-3s^2)\over \beta^2t}\nonumber \\
&&-{3(4t+s)^2\over 2\beta^4t}\biggr)\text{C}_0^{W,W}
\nonumber \\&&
+{32\pi^2\over 3}\biggl({2(t+2s)\over m_W^2}-{3t+2s-4m_W^2\over t}
-{t(t+s)\over m_W^4}\biggr),
\end{eqnarray}
\begin{eqnarray}
J_u^{1}(s,t)
&=&
-{128(t^2+2st+2s^2)\over m_W^2}
-{16(t^2-21st-26m_W^2t+34m_W^2s+17m_W^4)\over t}\nonumber\\
&& +{64st(t+s)\over m_W^4}+{32s^2\over t-m_W^2}
\nonumber \\
&&
+\biggl(16(t-5s+2m_W^2)-{48m_W^2(2s+m_W^2)\over t}
+{64s(2t+s)\over t-m_W^2}
\nonumber \\
&&
-{32s^2t\over (t-m_W^2)^2}\biggr)\ln\biggl({-t\over m_W^2}\biggr)
\nonumber \\
&&
+
\biggl(
{16(4t+s)\over\beta^2}
-16(3t-2s)+{48m_W^2(2t-2s-m_W^2)\over t}\biggr)
\ln\biggl({s\over m_W^2}\biggr)
\nonumber \\
&&+16s\biggl(t(2s+u)-2m_W^2(2s+m_W^2)\biggr)\text{D}_0^{W,W}\nonumber\\ &&
+32(t-m_W^2)\biggl({2m_W^2(2s+m_W^2)\over t}-2s-u\biggr)\text{C}^{W,t}_0
\nonumber \\ &&
+\biggl(32st-12s^2+32m_W^4-16m_W^2(2t+7s)-{4s(4t+s)\over\beta^2}\biggr)\text{C}_0^{W,W}
\nonumber \\
&&+{32\pi^2\over 3}\biggl({2(t^2+2st+2s^2)\over m_W^2}
-{st(t+s)\over m_W^4}\nonumber\\
&&-{2m_W^2(2t-2s-m_W^2)\over t}
-t-4s\biggr),
\end{eqnarray}
\begin{eqnarray}
K_u^{1}(s,t)&=& 16\bigg\{12t^2+20st-24m_W^2t+17s^2-4m_W^2s+12m_W^4+
{s^2t(t+s)\over m_W^4}
\nonumber \\
 &&
-{2s(2t^2+3st+2s^2)\over m_W^2}\bigg\}\left(2-{\pi^2\over 3}\right),
\end{eqnarray}
with $F^1_d(s,t)=F^1_u(s,u)$, $J^1_d(s,t)=-J^2_u(s,u)$, and $K^1_d(s,t)=K^1_u(s,u)$.
\bibliography{wz}

\begin{thebibliography}{43}
\expandafter\ifx\csname natexlab\endcsname\relax\def\natexlab#1{#1}\fi
\expandafter\ifx\csname bibnamefont\endcsname\relax
  \def\bibnamefont#1{#1}\fi
\expandafter\ifx\csname bibfnamefont\endcsname\relax
  \def\bibfnamefont#1{#1}\fi
\expandafter\ifx\csname citenamefont\endcsname\relax
  \def\citenamefont#1{#1}\fi
\expandafter\ifx\csname url\endcsname\relax
  \def\url#1{\texttt{#1}}\fi
\expandafter\ifx\csname urlprefix\endcsname\relax\def\urlprefix{URL }\fi
\providecommand{\bibinfo}[2]{#2}
\providecommand{\eprint}[2][]{\url{#2}}

\bibitem[{\citenamefont{Aaltonen et~al.}(2010)}]{Aaltonen:2009aa}
\bibinfo{author}{\bibfnamefont{T.}~\bibnamefont{Aaltonen}} \bibnamefont{et~al.}
  (\bibinfo{collaboration}{CDF Collaboration}),
  \bibinfo{journal}{Phys.Rev.Lett.} \textbf{\bibinfo{volume}{104}},
  \bibinfo{pages}{201801} (\bibinfo{year}{2010}), \eprint{0912.4500}.

\bibitem[{\citenamefont{Aaltonen et~al.}(2012)}]{CDF:2011ab}
\bibinfo{author}{\bibfnamefont{T.}~\bibnamefont{Aaltonen}} \bibnamefont{et~al.}
  (\bibinfo{collaboration}{CDF Collaboration}),
  \bibinfo{journal}{Phys.Rev.Lett.} \textbf{\bibinfo{volume}{108}},
  \bibinfo{pages}{101801} (\bibinfo{year}{2012}), \eprint{1112.2978}.

\bibitem[{\citenamefont{Abazov et~al.}(2013{\natexlab{a}})}]{Abazov:2013opa}
\bibinfo{author}{\bibfnamefont{V.~M.} \bibnamefont{Abazov}}
  \bibnamefont{et~al.} (\bibinfo{collaboration}{D0 Collaboration})
  (\bibinfo{year}{2013}{\natexlab{a}}), \eprint{1305.1258}.

\bibitem[{\citenamefont{Abazov et~al.}(2012)}]{Abazov:2012cj}
\bibinfo{author}{\bibfnamefont{V.~M.} \bibnamefont{Abazov}}
  \bibnamefont{et~al.} (\bibinfo{collaboration}{D0 Collaboration}),
  \bibinfo{journal}{Phys.Rev.} \textbf{\bibinfo{volume}{D85}},
  \bibinfo{pages}{112005} (\bibinfo{year}{2012}), \eprint{1201.5652}.

\bibitem[{\citenamefont{Abazov et~al.}(2013{\natexlab{b}})}]{D0:2013rca}
\bibinfo{author}{\bibfnamefont{V.~M.} \bibnamefont{Abazov}}
  \bibnamefont{et~al.} (\bibinfo{collaboration}{D0 Collaboration})
  (\bibinfo{year}{2013}{\natexlab{b}}), \eprint{1304.5422}.

\bibitem[{\citenamefont{Aad et~al.}(2012{\natexlab{a}})}]{Aad:2011xj}
\bibinfo{author}{\bibfnamefont{G.}~\bibnamefont{Aad}} \bibnamefont{et~al.}
  (\bibinfo{collaboration}{ATLAS Collaboration}),
  \bibinfo{journal}{Phys.Rev.Lett.} \textbf{\bibinfo{volume}{108}},
  \bibinfo{pages}{041804} (\bibinfo{year}{2012}{\natexlab{a}}),
  \eprint{1110.5016}.

\bibitem[{\citenamefont{Chatrchyan et~al.}(2013)}]{Chatrchyan:2012sga}
\bibinfo{author}{\bibfnamefont{S.}~\bibnamefont{Chatrchyan}}
  \bibnamefont{et~al.} (\bibinfo{collaboration}{CMS Collaboration}),
  \bibinfo{journal}{JHEP} \textbf{\bibinfo{volume}{1301}}, \bibinfo{pages}{063}
  (\bibinfo{year}{2013}), \eprint{1211.4890}.

\bibitem[{\citenamefont{Aad et~al.}(2011)}]{Aad:2011kk}
\bibinfo{author}{\bibfnamefont{G.}~\bibnamefont{Aad}} \bibnamefont{et~al.}
  (\bibinfo{collaboration}{ATLAS Collaboration}),
  \bibinfo{journal}{Phys.Rev.Lett.} \textbf{\bibinfo{volume}{107}},
  \bibinfo{pages}{041802} (\bibinfo{year}{2011}), \eprint{1104.5225}.

\bibitem[{\citenamefont{Aad et~al.}(2012{\natexlab{b}})}]{Aad:2012twa}
\bibinfo{author}{\bibfnamefont{G.}~\bibnamefont{Aad}} \bibnamefont{et~al.}
  (\bibinfo{collaboration}{ATLAS Collaboration}),
  \bibinfo{journal}{Eur.Phys.J.} \textbf{\bibinfo{volume}{C72}},
  \bibinfo{pages}{2173} (\bibinfo{year}{2012}{\natexlab{b}}),
  \eprint{1208.1390}.

\bibitem[{\citenamefont{Aad et~al.}(2012{\natexlab{c}})}]{Aad:2011cx}
\bibinfo{author}{\bibfnamefont{G.}~\bibnamefont{Aad}} \bibnamefont{et~al.}
  (\bibinfo{collaboration}{ATLAS Collaboration}), \bibinfo{journal}{Phys.Lett.}
  \textbf{\bibinfo{volume}{B709}}, \bibinfo{pages}{341}
  (\bibinfo{year}{2012}{\natexlab{c}}), \eprint{1111.5570}.

\bibitem[{\citenamefont{Aad et~al.}(2013)}]{ATLAS:2012mec}
\bibinfo{author}{\bibfnamefont{G.}~\bibnamefont{Aad}} \bibnamefont{et~al.}
  (\bibinfo{collaboration}{ATLAS Collaboration}), \bibinfo{journal}{Phys.Rev.}
  \textbf{\bibinfo{volume}{D87}}, \bibinfo{pages}{112001}
  (\bibinfo{year}{2013}), \eprint{1210.2979}.

\bibitem[{CMS(2013)}]{CMS-PAS-SMP-13-005}
\bibinfo{type}{Tech. Rep.} \bibinfo{number}{CMS-PAS-SMP-13-005},
  \bibinfo{institution}{CERN}, \bibinfo{address}{Geneva}
  (\bibinfo{year}{2013}).

\bibitem[{\citenamefont{Frixione}(1993)}]{Frixione:1993yp}
\bibinfo{author}{\bibfnamefont{S.}~\bibnamefont{Frixione}},
  \bibinfo{journal}{Nucl.Phys.} \textbf{\bibinfo{volume}{B410}},
  \bibinfo{pages}{280} (\bibinfo{year}{1993}).

\bibitem[{\citenamefont{Ohnemus and Owens}(1991)}]{PhysRevD.43.3626}
\bibinfo{author}{\bibfnamefont{J.}~\bibnamefont{Ohnemus}} \bibnamefont{and}
  \bibinfo{author}{\bibfnamefont{J.~F.} \bibnamefont{Owens}},
  \bibinfo{journal}{Phys. Rev. D} \textbf{\bibinfo{volume}{43}},
  \bibinfo{pages}{3626} (\bibinfo{year}{1991}).

\bibitem[{\citenamefont{Mele et~al.}(1991)\citenamefont{Mele, Nason, and
  Ridolfi}}]{Mele1991409}
\bibinfo{author}{\bibfnamefont{B.}~\bibnamefont{Mele}},
  \bibinfo{author}{\bibfnamefont{P.}~\bibnamefont{Nason}}, \bibnamefont{and}
  \bibinfo{author}{\bibfnamefont{G.}~\bibnamefont{Ridolfi}},
  \bibinfo{journal}{Nuclear Physics B} \textbf{\bibinfo{volume}{357}},
  \bibinfo{pages}{409 } (\bibinfo{year}{1991}), ISSN \bibinfo{issn}{0550-3213}.

\bibitem[{\citenamefont{Dixon et~al.}(1999)\citenamefont{Dixon, Kunszt, and
  Signer}}]{PhysRevD.60.114037}
\bibinfo{author}{\bibfnamefont{L.}~\bibnamefont{Dixon}},
  \bibinfo{author}{\bibfnamefont{Z.}~\bibnamefont{Kunszt}}, \bibnamefont{and}
  \bibinfo{author}{\bibfnamefont{A.}~\bibnamefont{Signer}},
  \bibinfo{journal}{Phys. Rev. D} \textbf{\bibinfo{volume}{60}},
  \bibinfo{pages}{114037} (\bibinfo{year}{1999}).

\bibitem[{\citenamefont{Campbell and Ellis}(1999)}]{PhysRevD.60.113006}
\bibinfo{author}{\bibfnamefont{J.~M.} \bibnamefont{Campbell}} \bibnamefont{and}
  \bibinfo{author}{\bibfnamefont{R.~K.} \bibnamefont{Ellis}},
  \bibinfo{journal}{Phys. Rev. D} \textbf{\bibinfo{volume}{60}},
  \bibinfo{pages}{113006} (\bibinfo{year}{1999}).

\bibitem[{\citenamefont{Dixon et~al.}(1998)\citenamefont{Dixon, Kunszt, and
  Signer}}]{Dixon19983}
\bibinfo{author}{\bibfnamefont{L.}~\bibnamefont{Dixon}},
  \bibinfo{author}{\bibfnamefont{Z.}~\bibnamefont{Kunszt}}, \bibnamefont{and}
  \bibinfo{author}{\bibfnamefont{A.}~\bibnamefont{Signer}},
  \bibinfo{journal}{Nuclear Physics B} \textbf{\bibinfo{volume}{531}},
  \bibinfo{pages}{3 } (\bibinfo{year}{1998}), ISSN \bibinfo{issn}{0550-3213}.

\bibitem[{\citenamefont{Chachamis et~al.}(2008)\citenamefont{Chachamis, Czakon,
  and Eiras}}]{Chachamis2008a}
\bibinfo{author}{\bibfnamefont{G.}~\bibnamefont{Chachamis}},
  \bibinfo{author}{\bibfnamefont{M.}~\bibnamefont{Czakon}}, \bibnamefont{and}
  \bibinfo{author}{\bibfnamefont{D.}~\bibnamefont{Eiras}},
  \bibinfo{journal}{Journal of High Energy Physics}
  \textbf{\bibinfo{volume}{2008}}, \bibinfo{pages}{22} (\bibinfo{year}{2008}),
  ISSN \bibinfo{issn}{1029-8479}, \eprint{0802.4028}.

\bibitem[{\citenamefont{Dawson et~al.}(2013)\citenamefont{Dawson, Lewis, and
  Zeng}}]{Dawson:2013lya}
\bibinfo{author}{\bibfnamefont{S.}~\bibnamefont{Dawson}},
  \bibinfo{author}{\bibfnamefont{I.~M.} \bibnamefont{Lewis}}, \bibnamefont{and}
  \bibinfo{author}{\bibfnamefont{M.}~\bibnamefont{Zeng}}
  (\bibinfo{year}{2013}), \eprint{1307.3249}.

\bibitem[{\citenamefont{Campanario and Sapeta}(2012)}]{Campanario:2012fk}
\bibinfo{author}{\bibfnamefont{F.}~\bibnamefont{Campanario}} \bibnamefont{and}
  \bibinfo{author}{\bibfnamefont{S.}~\bibnamefont{Sapeta}},
  \bibinfo{journal}{Phys.Lett.} \textbf{\bibinfo{volume}{B718}},
  \bibinfo{pages}{100} (\bibinfo{year}{2012}), \eprint{1209.4595}.

\bibitem[{\citenamefont{Bauer et~al.}(2001)\citenamefont{Bauer, Fleming,
  Pirjol, and Stewart}}]{Bauer2001}
\bibinfo{author}{\bibfnamefont{C.~W.} \bibnamefont{Bauer}},
  \bibinfo{author}{\bibfnamefont{S.}~\bibnamefont{Fleming}},
  \bibinfo{author}{\bibfnamefont{D.}~\bibnamefont{Pirjol}}, \bibnamefont{and}
  \bibinfo{author}{\bibfnamefont{I.~W.} \bibnamefont{Stewart}},
  \bibinfo{journal}{Physical Review D} \textbf{\bibinfo{volume}{63}},
  \bibinfo{pages}{114020} (\bibinfo{year}{2001}), ISSN
  \bibinfo{issn}{0556-2821}, \eprint{0011336}.

\bibitem[{\citenamefont{Bauer et~al.}(2002)\citenamefont{Bauer, Pirjol, and
  Stewart}}]{Bauer2002}
\bibinfo{author}{\bibfnamefont{C.~W.} \bibnamefont{Bauer}},
  \bibinfo{author}{\bibfnamefont{D.}~\bibnamefont{Pirjol}}, \bibnamefont{and}
  \bibinfo{author}{\bibfnamefont{I.~W.} \bibnamefont{Stewart}},
  \bibinfo{journal}{Physical Review D} \textbf{\bibinfo{volume}{65}},
  \bibinfo{pages}{054022} (\bibinfo{year}{2002}), ISSN
  \bibinfo{issn}{0556-2821}, \eprint{0109045}.

\bibitem[{\citenamefont{Beneke et~al.}(2002)\citenamefont{Beneke, Chapovsky,
  Diehl, and Feldmann}}]{Beneke2002}
\bibinfo{author}{\bibfnamefont{M.}~\bibnamefont{Beneke}},
  \bibinfo{author}{\bibfnamefont{A.}~\bibnamefont{Chapovsky}},
  \bibinfo{author}{\bibfnamefont{M.}~\bibnamefont{Diehl}}, \bibnamefont{and}
  \bibinfo{author}{\bibfnamefont{T.}~\bibnamefont{Feldmann}},
  \bibinfo{journal}{Nuclear Physics B} \textbf{\bibinfo{volume}{643}},
  \bibinfo{pages}{431} (\bibinfo{year}{2002}), ISSN \bibinfo{issn}{05503213},
  \eprint{0206152}.

\bibitem[{\citenamefont{Gao et~al.}(2005)\citenamefont{Gao, Li, and
  Liu}}]{Gao:2005iu}
\bibinfo{author}{\bibfnamefont{Y.}~\bibnamefont{Gao}},
  \bibinfo{author}{\bibfnamefont{C.~S.} \bibnamefont{Li}}, \bibnamefont{and}
  \bibinfo{author}{\bibfnamefont{J.~J.} \bibnamefont{Liu}},
  \bibinfo{journal}{Phys.Rev.} \textbf{\bibinfo{volume}{D72}},
  \bibinfo{pages}{114020} (\bibinfo{year}{2005}), \eprint{hep-ph/0501229}.

\bibitem[{\citenamefont{Idilbi et~al.}(2005)\citenamefont{Idilbi, Ji, and
  Yuan}}]{Idilbi:2005er}
\bibinfo{author}{\bibfnamefont{A.}~\bibnamefont{Idilbi}},
  \bibinfo{author}{\bibfnamefont{X.-d.} \bibnamefont{Ji}}, \bibnamefont{and}
  \bibinfo{author}{\bibfnamefont{F.}~\bibnamefont{Yuan}},
  \bibinfo{journal}{Phys.Lett.} \textbf{\bibinfo{volume}{B625}},
  \bibinfo{pages}{253} (\bibinfo{year}{2005}), \eprint{hep-ph/0507196}.

\bibitem[{\citenamefont{Becher and
  Neubert}(2011{\natexlab{a}})}]{Becher:2010tm}
\bibinfo{author}{\bibfnamefont{T.}~\bibnamefont{Becher}} \bibnamefont{and}
  \bibinfo{author}{\bibfnamefont{M.}~\bibnamefont{Neubert}},
  \bibinfo{journal}{Eur.Phys.J.} \textbf{\bibinfo{volume}{C71}},
  \bibinfo{pages}{1665} (\bibinfo{year}{2011}{\natexlab{a}}),
  \eprint{1007.4005}.

\bibitem[{\citenamefont{Becher et~al.}(2012{\natexlab{a}})\citenamefont{Becher,
  Neubert, and Wilhelm}}]{Becher:2011xn}
\bibinfo{author}{\bibfnamefont{T.}~\bibnamefont{Becher}},
  \bibinfo{author}{\bibfnamefont{M.}~\bibnamefont{Neubert}}, \bibnamefont{and}
  \bibinfo{author}{\bibfnamefont{D.}~\bibnamefont{Wilhelm}},
  \bibinfo{journal}{JHEP} \textbf{\bibinfo{volume}{1202}}, \bibinfo{pages}{124}
  (\bibinfo{year}{2012}{\natexlab{a}}), \eprint{1109.6027}.

\bibitem[{\citenamefont{Echevarria et~al.}(2012)\citenamefont{Echevarria,
  Idilbi, and Scimemi}}]{GarciaEchevarria:2011rb}
\bibinfo{author}{\bibfnamefont{M.~G.} \bibnamefont{Echevarria}},
  \bibinfo{author}{\bibfnamefont{A.}~\bibnamefont{Idilbi}}, \bibnamefont{and}
  \bibinfo{author}{\bibfnamefont{I.}~\bibnamefont{Scimemi}},
  \bibinfo{journal}{JHEP} \textbf{\bibinfo{volume}{1207}}, \bibinfo{pages}{002}
  (\bibinfo{year}{2012}), \eprint{1111.4996}.

\bibitem[{\citenamefont{Chiu et~al.}(2012)\citenamefont{Chiu, Jain, Neill, and
  Rothstein}}]{Chiu:2012ir}
\bibinfo{author}{\bibfnamefont{J.-Y.} \bibnamefont{Chiu}},
  \bibinfo{author}{\bibfnamefont{A.}~\bibnamefont{Jain}},
  \bibinfo{author}{\bibfnamefont{D.}~\bibnamefont{Neill}}, \bibnamefont{and}
  \bibinfo{author}{\bibfnamefont{I.~Z.} \bibnamefont{Rothstein}},
  \bibinfo{journal}{JHEP} \textbf{\bibinfo{volume}{1205}}, \bibinfo{pages}{084}
  (\bibinfo{year}{2012}), \eprint{1202.0814}.

\bibitem[{\citenamefont{Becher et~al.}(2013)\citenamefont{Becher, Neubert, and
  Wilhelm}}]{Becher:2012yn}
\bibinfo{author}{\bibfnamefont{T.}~\bibnamefont{Becher}},
  \bibinfo{author}{\bibfnamefont{M.}~\bibnamefont{Neubert}}, \bibnamefont{and}
  \bibinfo{author}{\bibfnamefont{D.}~\bibnamefont{Wilhelm}},
  \bibinfo{journal}{JHEP} \textbf{\bibinfo{volume}{1305}}, \bibinfo{pages}{110}
  (\bibinfo{year}{2013}), \eprint{1212.2621}.

\bibitem[{\citenamefont{Becher and Neubert}(2011{\natexlab{b}})}]{Becher2011}
\bibinfo{author}{\bibfnamefont{T.}~\bibnamefont{Becher}} \bibnamefont{and}
  \bibinfo{author}{\bibfnamefont{M.}~\bibnamefont{Neubert}},
  \bibinfo{journal}{The European Physical Journal C}
  \textbf{\bibinfo{volume}{71}}, \bibinfo{pages}{1665}
  (\bibinfo{year}{2011}{\natexlab{b}}), ISSN \bibinfo{issn}{1434-6044},
  \eprint{1007.4005}.

\bibitem[{\citenamefont{Becher et~al.}(2012{\natexlab{b}})\citenamefont{Becher,
  Neubert, and Wilhelm}}]{Becher2012}
\bibinfo{author}{\bibfnamefont{T.}~\bibnamefont{Becher}},
  \bibinfo{author}{\bibfnamefont{M.}~\bibnamefont{Neubert}}, \bibnamefont{and}
  \bibinfo{author}{\bibfnamefont{D.}~\bibnamefont{Wilhelm}},
  \bibinfo{journal}{Journal of High Energy Physics}
  \textbf{\bibinfo{volume}{2012}}, \bibinfo{pages}{124}
  (\bibinfo{year}{2012}{\natexlab{b}}), ISSN \bibinfo{issn}{1029-8479},
  \eprint{1109.6027}.

\bibitem[{\citenamefont{Grazzini}(2005)}]{Grazzini2005a}
\bibinfo{author}{\bibfnamefont{M.}~\bibnamefont{Grazzini}},
  \bibinfo{journal}{Journal of High Energy Physics}
  \textbf{\bibinfo{volume}{2006}}, \bibinfo{pages}{15} (\bibinfo{year}{2005}),
  ISSN \bibinfo{issn}{1029-8479}, \eprint{0510337}.

\bibitem[{\citenamefont{Frederix and Grazzini}(2008)}]{Frederix2008}
\bibinfo{author}{\bibfnamefont{R.}~\bibnamefont{Frederix}} \bibnamefont{and}
  \bibinfo{author}{\bibfnamefont{M.}~\bibnamefont{Grazzini}},
  \bibinfo{journal}{Physics Letters B} \textbf{\bibinfo{volume}{662}},
  \bibinfo{pages}{353} (\bibinfo{year}{2008}), ISSN \bibinfo{issn}{03702693},
  \eprint{0801.2229}.

\bibitem[{\citenamefont{Bal\'{a}zs and Yuan}(1999)}]{Balazs1999}
\bibinfo{author}{\bibfnamefont{C.}~\bibnamefont{Bal\'{a}zs}} \bibnamefont{and}
  \bibinfo{author}{\bibfnamefont{C.-P.} \bibnamefont{Yuan}},
  \bibinfo{journal}{Physical Review D} \textbf{\bibinfo{volume}{59}},
  \bibinfo{pages}{114007} (\bibinfo{year}{1999}), ISSN
  \bibinfo{issn}{0556-2821}, \eprint{9810319v4}.

\bibitem[{\citenamefont{Bal\'{a}zs and Yuan}(2001)}]{Balazs2001a}
\bibinfo{author}{\bibfnamefont{C.}~\bibnamefont{Bal\'{a}zs}} \bibnamefont{and}
  \bibinfo{author}{\bibfnamefont{C.-P.} \bibnamefont{Yuan}},
  \bibinfo{journal}{Physical Review D} \textbf{\bibinfo{volume}{63}},
  \bibinfo{pages}{059902} (\bibinfo{year}{2001}), ISSN
  \bibinfo{issn}{0556-2821}.

\bibitem[{\citenamefont{Li et~al.}(2013)\citenamefont{Li, Li, Shao, Yang, and
  Zhu}}]{Li:2013mia}
\bibinfo{author}{\bibfnamefont{H.~T.} \bibnamefont{Li}},
  \bibinfo{author}{\bibfnamefont{C.~S.} \bibnamefont{Li}},
  \bibinfo{author}{\bibfnamefont{D.~Y.} \bibnamefont{Shao}},
  \bibinfo{author}{\bibfnamefont{L.~L.} \bibnamefont{Yang}}, \bibnamefont{and}
  \bibinfo{author}{\bibfnamefont{H.~X.} \bibnamefont{Zhu}}
  (\bibinfo{year}{2013}), \eprint{1307.2464}.

\bibitem[{\citenamefont{Frixione et~al.}(1992)\citenamefont{Frixione, Nason,
  and Ridolfi}}]{Frixione:1992pj}
\bibinfo{author}{\bibfnamefont{S.}~\bibnamefont{Frixione}},
  \bibinfo{author}{\bibfnamefont{P.}~\bibnamefont{Nason}}, \bibnamefont{and}
  \bibinfo{author}{\bibfnamefont{G.}~\bibnamefont{Ridolfi}},
  \bibinfo{journal}{Nucl.Phys.} \textbf{\bibinfo{volume}{B383}},
  \bibinfo{pages}{3} (\bibinfo{year}{1992}).

\bibitem[{\citenamefont{Becher et~al.}(2008)\citenamefont{Becher, Neubert, and
  Xu}}]{Becher:2007ty}
\bibinfo{author}{\bibfnamefont{T.}~\bibnamefont{Becher}},
  \bibinfo{author}{\bibfnamefont{M.}~\bibnamefont{Neubert}}, \bibnamefont{and}
  \bibinfo{author}{\bibfnamefont{G.}~\bibnamefont{Xu}}, \bibinfo{journal}{JHEP}
  \textbf{\bibinfo{volume}{0807}}, \bibinfo{pages}{030} (\bibinfo{year}{2008}),
  \eprint{0710.0680}.

\bibitem[{\citenamefont{Becher and Bell}(2012)}]{Becher:2011dz}
\bibinfo{author}{\bibfnamefont{T.}~\bibnamefont{Becher}} \bibnamefont{and}
  \bibinfo{author}{\bibfnamefont{G.}~\bibnamefont{Bell}},
  \bibinfo{journal}{Phys.Lett.} \textbf{\bibinfo{volume}{B713}},
  \bibinfo{pages}{41} (\bibinfo{year}{2012}), \eprint{1112.3907}.

\bibitem[{\citenamefont{Beringer et~al.}(2012)}]{Beringer:1900zz}
\bibinfo{author}{\bibfnamefont{J.}~\bibnamefont{Beringer}} \bibnamefont{et~al.}
  (\bibinfo{collaboration}{Particle Data Group}), \bibinfo{journal}{Phys.Rev.}
  \textbf{\bibinfo{volume}{D86}}, \bibinfo{pages}{010001}
  (\bibinfo{year}{2012}).

\bibitem[{\citenamefont{Campbell and Ellis}(2000)}]{Campbell2000}
\bibinfo{author}{\bibfnamefont{J.~M.} \bibnamefont{Campbell}} \bibnamefont{and}
  \bibinfo{author}{\bibfnamefont{R.~K.} \bibnamefont{Ellis}},
  \bibinfo{journal}{Physical Review D} \textbf{\bibinfo{volume}{62}},
  \bibinfo{pages}{33} (\bibinfo{year}{2000}), ISSN \bibinfo{issn}{0556-2821},
  \eprint{0006304}.

\end{thebibliography}
\end{document}